\documentclass[%
 reprint,
 amsmath,amssymb,
 pra,
floatfix,
]{revtex4-1}

\usepackage{graphicx}
\usepackage{dcolumn}
\usepackage{bm}

\newcommand{\arccosh}{\,{\rm arccosh}}
\newcommand{\arcsinh}{\,{\rm arcsinh}}
\usepackage{xcolor,soul}

\begin{document}

\title{Analysis of parameter combinations for optimal soliton microcomb
    generation efficiency in simple single-cavity scheme}
\author{Nikita~M.~ Kondratiev\textsuperscript{1}}\email{nikita.kondratyev@tii.ae}
\author{Valery~E.~ Lobanov\textsuperscript{2}}
\author{Nikita~Yu.~ Dmitriev\textsuperscript{2}}
\author{Steevy~J.~ Cordette\textsuperscript{1}}
\author{Igor~A.~Bilenko\textsuperscript{2,3}}
\affiliation{\textsuperscript{1}Directed Energy Research Centre, Technology Innovation Institute, Abu Dhabi, United Arab Emirates}
\affiliation{\textsuperscript{2}Russian Quantum Center, 143026, Skolkovo, Russia}
\affiliation{\textsuperscript{3}Faculty of Physics, Lomonosov Moscow State University, 119991, Moscow, Russia}

\date{\today}

\begin{abstract}
Dissipative Kerr solitons generated in high-Q optical microresonators provide unique opportunities for different up-to-date applications. Increasing the generation efficiency of such signals is a problem of paramount importance. We performed a comprehensive {analytical and numerical analysis in simple single-cavity scheme. It was revealed that in order to obtain high pump-to-comb conversion efficiency such parameters as coupling rate, pump amplitude, detuning and microresonator second-order dispersion should not be considered individually, but only in the aggregate. The dependence of the optimal coupling rate on the pump power was shown, in addition to the trade-off relations balancing the efficiency versus the number of comb lines. Combining analytical predictions and numerical simulations, we found optimal conditions for  maximal pump-to-comb conversion efficiency (up to 100\%) in the cases of free-running and self-injection-locked pump lasers. The discrepancy between numerical and analytical solutions and methods to increase the total comb power were also discussed.}
\end{abstract}

\maketitle


\section{INTRODUCTION}
\label{sec:intro}  
Microresonator-based frequency combs or microcombs \cite{Kippenberg2011,Chembo:2016,PASQUAZI20181,Gaeta2019,Kovach:20,Nie:2022} are promising tools for numerous applications including Time-Frequency Metrology, Optical Frequency Synthesis, Spectroscopy, Communication, and  Quantum Technologies \cite{Suh600,Kues2019, Riemensberger2020, Marin-Palomo2017,Suh2019,Kippenbergeaan8083,Sun:23}. Coherent frequency combs, or dissipative solitons in temporal representation, are of particular interest. Unfortunately, the most accessible coherent frequency combs in the form of bright dissipative Kerr solitons \cite{Kippenbergeaan8083} are realisable in anomalous group velocity dispersion (GVD) regime, and suffer from a low pump-to-comb conversion efficiency which had long been commonly known to be no more than 10\% \cite{Kippenbergeaan8083,Karpov2019}. To overcome this limitation, special feedback schemes \cite{Xue2019,Boggio2022}, normal dispersion usage \cite{Xue:17,KondratievNum:20,Kim:19,Ji:23} and pulsed pumping \cite{Li:22} have been suggested. 

However, recent studies demonstrated that it is possible to reach the values up to 40\% \cite{dmitriev2021hybrid,jang2021conversion} even without any special schemes, showing that this problem has not been fully investigated, and no fundamental limit of the conversion efficiency has been clearly defined. Although previous studies concerned the dependence of the pump conversion efficiency on the microresonator free spectral range (FSR) \cite{jang2021conversion}, or the input power and the coupling rate \cite{Bao:14,Gartner2019}, those dependencies were usually treated independently, and only general trends were shown. In reality, both FSR and coupling rate influence the threshold power -- the former through the mode volume, the latter through the loaded quality factor. Furthermore, restrictions should be applied on the coupling rate and the pump power in order to keep the system above the comb generation threshold. This suggests a nontrivial optimization problem that has never been addressed. {We also note that all previous works relied on the simple sech-squared form of the soliton \cite{herr2014temporal}, that is known to be a good approximation only for rather large values of pump detuning and amplitude.}

{In this paper we use simple analytical approach to identify the regions of high soliton microcomb generation efficiency considering the combination of significant parameters of the system "laser-microresonator" and then verify and amend obtained results numerically. We demonstrate how the pump power, pump frequency, laser-to-microresonator coupling rate, and second-order dispersion of microresonator interplay to allow up to a 100\%-high conversion without any additional schemes, but at the cost of the comb line number. Effective methods to increase the total comb power are also discussed. The analysis is performed for the cases of the free-running \cite{herr2014temporal} and self-injection-locked \cite{Pavlov2018,Raja2019,shen2019} pump lasers. We also show that moderate backscattering that is necessary for self-injection locking does not suppress high conversion. } 

\section{CONVERSION EFFICIENCY}

We begin by defining the conversion efficiency figure of merit.
Two main definitions of comb generation efficiency can be found in literature: the pump to total comb (pump to soliton) generation efficiency $\eta_{\rm p2s}$ \cite{jang2021conversion,Bao:14,Yi:16} and pump to comb (pump to sidebands) generation efficiency $\eta_{\rm p2c}$ \cite{Xue:17,Boggio2022}. The first variant is more straightforward from a theoretical point of view, as in analytical approximation the field can be naturally decomposed into a soliton part and a background part. In this temporal representation, the background is just a constant field (stationary single-mode solution) $a_{\rm st}$, whereas the soliton part is a field having a train of hyperbolic-secant pulses each having the spectral components:
\begin{align}
\label{comblines}
    a_\mu&=\sqrt{\frac{d_2}{2}}\,{\rm sech}\left(\frac{\pi\mu}{2}\sqrt{\frac{d_2}{\zeta}}\right)e^{i\psi^{\rm sol}}, \\
    \psi^{\rm sol}&=\arctan\frac{\sqrt{2\pi^2f^2\zeta-16\zeta^2}}{4\zeta},
\end{align}
where $d_2=D_2/\kappa$ is the normalized GVD coefficient (assuming the microresonator eigenfrequencies  $\omega_\mu=\omega_0+D_1\mu+D_2\mu^2/2$, where $\omega_0$ is the eigenfrequency closest to the pump, $D_1$ is the microresonator FSR and $\mu$ is the relative mode number from the pumped mode; high-order dispersion terms are neglected), $\kappa$ is the pumped mode linewidth, $\zeta=2(\omega_0 - \omega_{\rm gen})/\kappa$ is the normalized pump frequency detuning (with $\omega_{\rm gen}$ the laser generation frequency) and $f=\sqrt{P/P_{\rm th}}$ is the normalized pump amplitude equal to the ratio of pump power to the nonlinear threshold power
\begin{align}
\label{Pthreshold0}
P_{\rm th}=\frac{\omega_0 n_g^2V_{\rm eff}}{8cn_2Q^2\eta},
\end{align}
where 
$Q=\omega_0 / \kappa$ is the loaded quality factor,
$n_2$ is the Kerr nonlinear refractive index,
$n_g$ is the microresonator mode group index,
$V_{\rm eff}$ is the microresonator effective mode volume. We should note that \eqref{comblines} is exact only for a Nonlinear Schr$\rm \ddot o$dinger Equation that is undumped undriven Lugiato-Lefever equation (LLE), being a limit case at $\zeta\gg1$ \cite{herr2014temporal}. This limitation is also in line with a natural demand for a soliton width to be smaller than the microresonator circumference $\sqrt{d_2/\zeta}\ll2\pi$. {As we see further, this brings limitations to the theories based on this formalism.}
The output average power of solitons can be easily calculated by summing the mode components of the hyperbolic-secant-squared shaped comb power \cite{dmitriev2021hybrid}, and recalculating the corresponding output. Using the transmittance coefficient from the ring resonator to the straight waveguide $T^2=\eta\kappa t_{\rm rt}$, where $t_{\rm rt}=n_{\rm eff}L/c$ is the round-trip time \cite{Gorodetsky:99}, we obtain
\begin{align}
\label{Pouttotal}
P_{\rm soliton}^{\rm out}=\frac{8\eta^2}{\pi}\sqrt{d_2\zeta}P_{\rm th},
\end{align}
where $\eta=\kappa_c/\kappa$ is the normalized coupling rate or coupling efficiency ($\eta=1/2$ for critical coupling),
$\kappa_c$ is the coupling rate.
Note that we consider anomalous GVD ($D_2>0$) and use the negative detuning definition, where the solitons exist at positive $\zeta$.
In this context, it is convenient to have a dimensional constant independent of the coupling efficiency. Therefore, we introduce the minimal threshold power $P_0=\frac{27}{32}\frac{n_g^2V_{\rm eff}\kappa_0^2}{cn_2\omega_0}$, where $\kappa_0=\kappa-\kappa_c$ is the microresonator intrinsic decay rate.
Then, the parametric instability threshold power is rewritten as follows:
\begin{align}
\label{Pthreshold}
P_{\rm th}=\frac{4P_0}{27}\frac{1}{\eta(1-\eta)^2},
\end{align}
The minimal threshold power $P_0$ is achieved at $\eta=1/3$, which corresponds to the undercoupling regime, while the threshold power at critical coupling is $1.2$ times higher. Remarkably, fulfilling the criterion $\eta=1/3$ coincides with the optimal coupling efficiency for the self-injection locking at weak backscattering \cite{Galiev2020}.
Then, dividing \eqref{Pouttotal} by the input power at the coupler $P$, we immediately get $\eta_{\rm p2s}$.

Experimentally, the soliton parameters are difficult to measure in the time domain because of the ultra-broad bandwidth required for their detection, while the optical power spectrum is usually easier to measure and enough to determine the average power of the soliton thanks to Parseval’s theorem. The spectral representation of the solitons results in a frequency comb with a uniform phase under a hyperbolic-secant envelope \eqref{comblines}, while the constant background field results in addition to the central line (or carrier) at the pumped frequency. Thus, it is not straightforward to separate the background from the solitons. So it is easier to omit the whole central peak, thus obtaining the pump to comb sidebands or simply pump to comb efficiency \cite{dmitriev2021hybrid}:
\begin{align}
\label{eff_p2c}
\eta_{\rm p2c}=\frac{4\eta^2}{f^2}\left(\frac{2}{\pi}\sqrt{d_2\zeta}-\frac{d_2}{2}\right)=\eta_{\rm p2s}-\frac{4\eta^2}{f^2}\frac{d_2}{2},
\end{align}
The first summand is effectively "pump to solitons" efficiency $\eta_{\rm p2s}$, and the subtracted term is the central line power of the hyperbolic-secant envelope. It can also be more informative for those who more interested in sideband power than in total comb power. The major difference between these two efficiencies is their behaviour with the increase of the soliton number \cite{dmitriev2021hybrid}. The "pump to solitons" naturally grows linearly as the soliton powers add up, while the second grows slower and saturates near the maximum soliton number $N^{\rm sol}_{\rm max}\approx\sqrt{1/d_2}$ set by the minimum distance at which solitons do not interact with each other \cite{Karpov2019}. Equation \eqref{eff_p2c} shows that both efficiencies go down as the square of the pump amplitude, suggesting it to be as low as possible.

\begin{figure}[t!]
\centering
\includegraphics[width=0.88\linewidth]{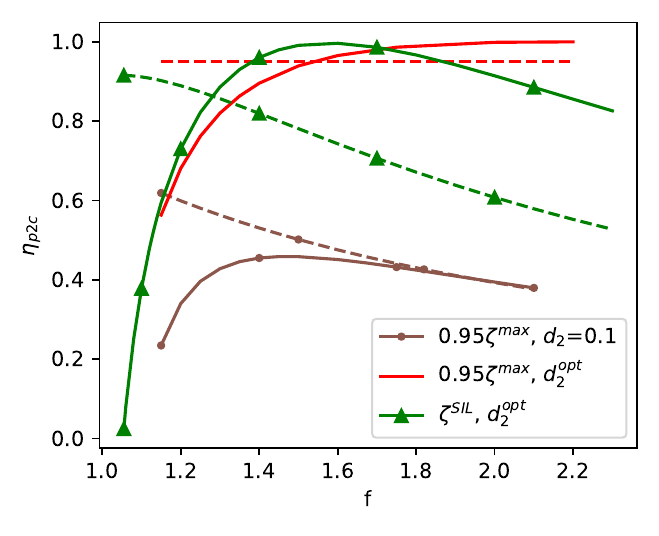}
\caption{Pump-to-comb conversion efficiency vs. normalized pump amplitude for different values of effective pump frequency detuning and normalized dispersion coefficient. Solid lines -- numerical modelling, dashed lines -- analytical expression \eqref{eff_p2c}. All quantities are plotted in dimensionless units.}
\label{analVSnum}
\end{figure}

\section{Analytical consideration}
In what following, we study the pump-to-comb efficiency as the most informative. Consider the parameters constituting the definition \eqref{eff_p2c}.
Generally, the detuning $\zeta$ is a free parameter. Obviously, it should be as large as possible for higher efficiency. However, a stabilisation of $\zeta$ is required to ensure the locking of the system to the desired regime despite the influence of various fluctuations \cite{herr2014temporal}. The easiest way to achieve this stabilisation is the self-injection locking (SIL) mechanism \cite{Kondratiev:17}, which locks the pump frequency to the microresonator eigenfrequency. As a result, we have two key ways to fix the range of the optimal detuning: (i) the maximum possible detuning enabling soliton generation (cut-off) $\zeta^{\rm max}=\pi^2f^2/8$ and (ii) the detuning provided by nonlinear SIL regime \cite{dmitriev2021hybrid,Kondratyev:22} $\zeta^0=2\sqrt{3}\sinh\left(\frac{1}{3}\arcsinh\left(\frac{3\sqrt{3}}{4}f^2\right)\right) $. 
We can see that $\zeta^0 \leq  \zeta^{\rm max}$ for $f>1.05$. For small pump powers, which are optimal to achieve high generation efficiency, those both boundary limits converge to $\zeta\approx1.38$ with $\zeta^{\rm max}$ hard (or impossible) to reach.

It is important to note, that both the SIL and maximal detunings do not depend on GVD coefficient. Thus, the pump-to-comb efficiency, \eqref{eff_p2c},  has a global maximum over the dispersion coefficient
\begin{align}
\label{eta_maxd2}
\eta_{\rm p2c}=\frac{8\eta^2}{f^2}\frac{\zeta}{\pi^2}\ \ {\rm at}\ \ d_2=\frac{4\zeta}{\pi^2},
\end{align}
while the pump-to-soliton efficiency $\eta_{\rm p2s}$ is monotonic over $d_2$. We can see that the global maximum of pump-to-comb efficiency $\eta_{\rm p2c}$ is 100 \% at $\eta=1$ regardless of the pump power (in soliton existance range) if we lock to the maximal detuning [see dashed line in Fig. \ref{analVSnum}] and 90.9 \% for the SIL state at near-threshold pump [see dashed line with triangles in Fig. \ref{analVSnum}]. Note that the  comb width normalised to units of the intermode distance $D_1$ at -3 dB level in this regime is $N_{\rm lines}=\frac{2}{\pi}\sqrt{\frac{\zeta}{d_2}}\times 2\arccosh\sqrt{2}\approx1.76$. This number effectively shows a number of comb line pairs around the pump above the -3 dB level. Here we come to the trade-off problem between the efficiency and the number of comb lines. 
The optimum described by \eqref{eta_maxd2} gives quite high dispersion values ($d_2\geq0.45$), revealing two challenges to be addressed. Firstly, the experimental realisation will require special resonator engineering \cite{Grudinin:15,FujiiTanabe2020,Wang:22,Lucas:22_arx,Zhang:22}. Until now in most reported experiments values $d_2<0.1$ were used as low dispersion allows for wider combs. 
Secondly, the validity of the analytical analysis as $d_2\geq1$ provides $N_{\rm sol}^{\rm max}<1$, coinciding with the state when the width of the hyperbolic-secant-squared soliton becomes comparable with the microresonator circumference. Although the soliton still exists for greater dispersion values, lower pump, and detunings (for SIL optimum), hyperbolic secant ceases to be a proper stationary solution, and the formula \eqref{Pouttotal} becomes inaccurate. So, we performed the numerical analysis of dissipative Kerr soliton properties using Lugiato-Lefever equation formalism \cite{Chembo2013,Godey2014}. This study showed that the expression \eqref{eff_p2c} is accurate only close to (but not exactly at) the cut-off detunings and $f>2$ [see Fig. \ref{analVSnum}, solid lines]. For SIL detuning, there is also a possibility to reach 100\% efficiency [see Fig. \ref{analVSnum}, lines with triangles], found numerically. More details are presented in Section 4.
{Note also that the desired high value of the second-order dispersion justifies the neglection of higher order dispersion terms that have strong impact in the regime of vanishing group velocity dispersion \cite{Bao:17_d3,TallaMbe:20,Anderson2022,Xiao2023,zhang2022microresonator,Ji:23}.}

\begin{figure}[t!]
\centering													
\includegraphics[width=0.93\linewidth]{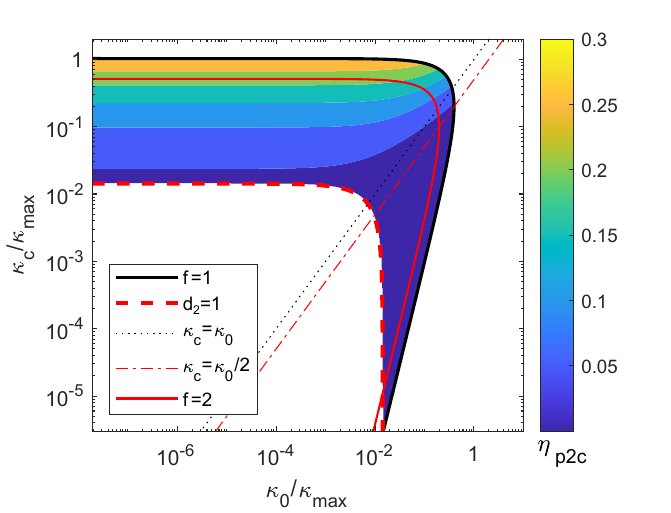}	 
\caption{Pump-to-comb conversion efficiency vs. coupling rate and internal losses calculated via \eqref{eff_p2c_kk_f1} for fixed pump power. 
$\zeta=\zeta^{\rm max}$ and 
$D_2/\kappa_{\rm max}=0.0143$.
The dotted and dash-dotted line correspond to the critical and minimal threshold coupling. Red (gray) solid curve shows $f=2$ (the isocontour for this is $\bar \kappa_0=(\bar\kappa_c^{1/3}-\bar\kappa_c)\kappa_{\rm max}/f$). All quantities are plotted in dimensionless units.}
\label{fig:eff(Q)}
\end{figure}

\begin{figure}[t]
\centering
\includegraphics[width=0.9\linewidth]{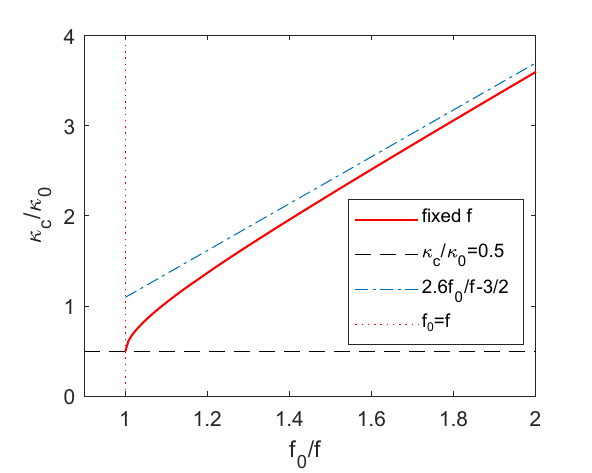}
\caption{Coupling efficiency needed to maintain fixed $f$ for the given pump power calculated from \eqref{kappacmin} (solid line). Dash-dotted line shows its asymptotic for high pump power. All quantities are plotted in dimensionless units.}
\label{kappac_f=1}
\end{figure}

To get more deep insight into the problem, we continue studying the equation \eqref{eff_p2c} for the near-cut-off detuning, where the analytic solution is still accurate. We have several restrictions on the parameters. First, the normalized pump amplitude should be greater than 1 for the soliton generation to be possible and then $f>2$ for the analytics to be valid. 
Secondly, the soliton should fit the ring $d_2<1$. This restriction also makes the number of lines $N_{\rm lines}>1$ (for $\zeta\approx1$), which is quite close to the natural demand for having more than one line pair in the comb around the pump. 
Last but not least, the coupling efficiency $\eta\in[0;1)$ by definition. Outstandingly, these restrictions are independent on the pump frequency detuning and conversion efficiency definition choice.

\begin{figure*}
\centering
\includegraphics[width=0.49\linewidth]{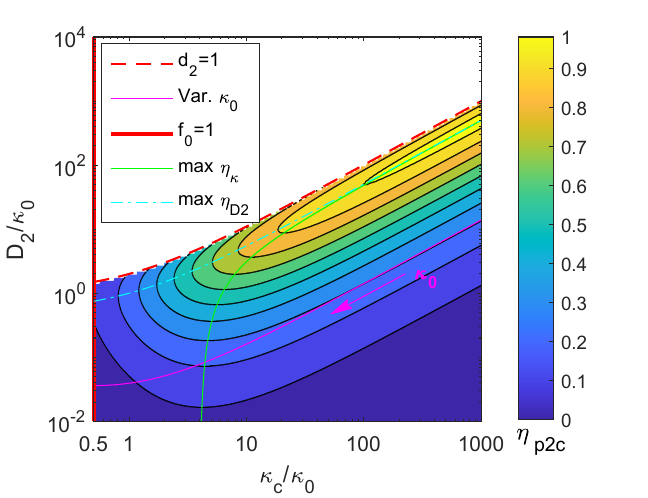}
\includegraphics[width=0.49\linewidth]{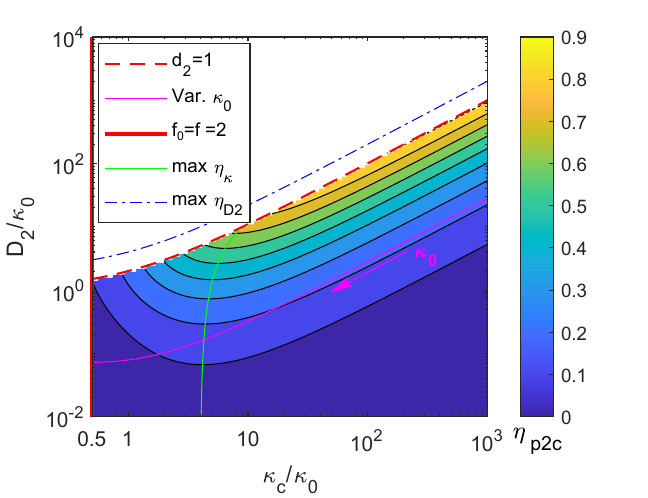}
\caption{
Pump-to-comb conversion efficiency vs. intrinsic-loss-normalised coupling $\kappa_c/\kappa_0$ and dispersion $D_2 / \kappa_0$ via \eqref{eff_p2c} and \eqref{kappacmin} for $\zeta=\zeta^{\rm max}$ and fixed $f$. Left panel shows $f=1$ and right panel -- $f=2$. The thick vertical line is minimal $P/P_0=1$, the dark (red) dashed line is $d_2=1$ and thin (pink) solid curve is the tuning of the intrinsic loss for 
$D_2/\kappa_{\rm max}=0.0143$.
Light (green) solid line is optimal coupling for fixed $D_2/\kappa_0$ and dash-dotted curve is the optimum \eqref{eta_maxd2}.
All quantities are plotted in dimensionless units.
}
\label{fig:eff(k)1.5}
\end{figure*}
\begin{figure*}[t!]
\centering
\includegraphics[width=0.47\linewidth]{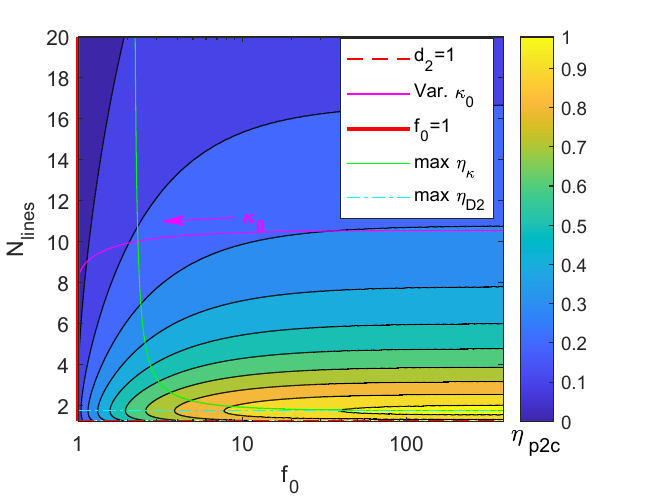}
\includegraphics[width=0.47\linewidth]{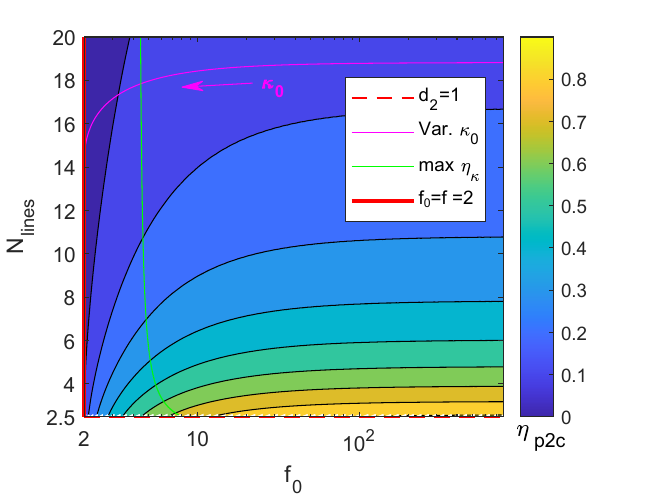}
\caption{Pump-to comb-efficiency vs. $f_0$ and comb lines number calculated via \eqref{eff_p2c} and \eqref{kappacmin}. Additional lines and parameters the same as in right panel of Fig. \ref{fig:eff(k)1.5}. All quantities are plotted in dimensionless units. 
}
\label{fig:eff(N)}
\end{figure*}

\subsection{Fixed pump power and dispersion}
{First consider the case of fixed pump power $P$ and second-order dispersion $D_2$.} Figure \ref{fig:eff(Q)} shows the dependence of $\eta_{\rm p2c}$ on the coupling rate $\kappa_c$ and internal losses $\kappa_0$ for the cut-off detuning and fixed other parameters (silicon nitride \cite{dmitriev2021hybrid}). We can see that for fixed pump power both $\kappa_c$ and $\kappa_0$ can not be chosen arbitrarily. Using \eqref{Pthreshold} and $f=1$ we introduce a maximal possible coupling rate for soliton to exist at given pump power:
\begin{equation}
 \kappa_{\rm max}=\sqrt{\frac{27P}{4P_0}}\kappa_0=\sqrt{\frac{8cn_2\omega_0P}{n_g^2V_{\rm eff}}}.   
\end{equation}
{For desired normalized pump amplitude $f$ at given pump power the value which bounds the coupling $\kappa_c$ is $\kappa_{\rm max}/f$ [see horizontal part of the upper black curve in Fig. \ref{fig:eff(Q)}] and for the maximal internal loss, bounding $\kappa_0$ is $\sqrt{4/27}\kappa_{\rm max}/f$ [see intersection of the upper black line and dash-dotted line in Fig. \ref{fig:eff(Q)}]. 
} Note that the $\kappa_{\rm max}$ actually does not depend on $\kappa_0$ as $P_0\propto \kappa_0^2$ and thus consists only of other trivial parameters and pump power.
To increase the level of generalisation, we normalise all rate coefficients to the maximal coupling rate at given $f$ and dispersion -- to the maximal coupling rare. Then for the pump to comb efficiency \eqref{eff_p2c}, we get
\begin{align}
\label{eff_p2c_kk_f1}
\eta_{\rm p2c}=\frac{4\bar\kappa_c}{\bar\kappa_0+\bar\kappa_c}\left(\sqrt{\bar d_2\bar\kappa_c}- \bar d_2(\bar\kappa_0+\bar\kappa_c) \right).
\end{align}
Here $\bar\kappa_c=\kappa_c{f}/\kappa_{\rm max}$, $\bar\kappa_0=\kappa_0{f}/\kappa_{\rm max}$, and $\bar d_2=D_2/2/\kappa_{\rm max}$. We also used $\zeta=\frac{\pi^2 f^2}{8}$ as we showed in previous section that the theory works better near cut-off. 

{Note again, that at fixed pump power, we can operate with different normalized pump amplitude $f$ by choosing different coupling and loss rate combinations, and get different efficiency [see Fig. \ref{fig:eff(Q)}, solid red (gray) curve $f=2$ for example].}
An important observation is that the maximum efficiency always lies on the boundary $f=1$ curve, confirming our previous speculations. Numerical study showed that the analytical solution is not accurate for low $f<2$, after which the efficiency drops.
So, for both definitions of the efficiency and both maximum detuning values we can use $f=2$ (e.g. the normalised pump should be minimised closer to modulational instability threshold) and strong overcoupling $\eta=1$ to maximise the conversion. Maximum value for the \eqref{eff_p2c_kk_f1}, found in $(\kappa_c,\kappa_0)\rightarrow(\kappa_{\rm max}/{f},0)$ (that is $\eta=1$ and $d_2=D_2/2/\kappa_{\rm max}$), can be written as $\eta_{\rm p2c}\approx4(\sqrt{\bar d_2/f}-\bar d_2/f)$. 
For typical 1 THz Si$_3$N$_4$ microresonator \cite{dmitriev2021hybrid},  the $P_0\approx9$ mW and $\kappa_0=188$ MHz. Then for the pump power $P\approx40$ mW, the maximal coupling rate is $\kappa_{\rm max}\approx 1$ GHz. Having $D_2\approx14.3$ MHz and $f=2$ we get $\eta_{\rm p2c}=22$\% for single-soliton comb.

\subsection{Fixed normalised pump amplitude}
{Now consider the case of fixed normalised pump $f$ and we optimise the dispersion and pump power or coupling. In previous sections, we saw that the normalised pump amplitude governs the comb generation regime, influencing strongly the efficiency and analitycs validity.}
After we fix $f$ the detuning values $\zeta^{\rm max}$ and $\zeta^0$ become fixed, and the expression \eqref{eff_p2c} has two free parameters: normalised dispersion coefficient $d_2$ and coupling coefficient $\eta$.
The last one is the easiest to be tuned. However, the first parameter also depends on the coupling, so it is better to extract it explicitly for further optimisation. Considering \eqref{Pthreshold}, the optimal pumping regime over $f$ imposes restrictions to the coupling
\begin{align}
\label{kappacmin}
\hat\kappa_c=\frac{\kappa_c}{\kappa_0}=3\frac{f_0}{f}\sin \left(\frac{1}{3}\arccos\frac{f}{f_0}+\frac{\pi}{6}\right)-1,
\end{align}
where we introduce normalized pump amplitude to minimal threshold $f_0=\sqrt{P/P_0}$ for simplicity. Formula \eqref{kappacmin} directly connects the required pump power with the optimal coupling level without using other parameters. This dependence is shown in Fig. \ref{kappac_f=1}. For large pump $f_0/f\gg3$ it tends to $\hat\kappa_c=3\frac{\sqrt{3}}{2}\frac{f_0}{f}-\frac{3}{2}$. Note, the minimal possible pump power $P=P_0f_x^2$ to maintain desired $f$ is obtained at subcritical coupling $\kappa_c=\kappa_0/2$ which is in agreement with \eqref{Pthreshold}. {Note also that for in this case it is more convenient to use rates normalised to the internal loss as $\kappa_{\rm max}$ is not fixed. Similarly, the minimal threshold power $P_0$ was not used in the previous subsection, being fixed only now.}

Combining the expressions \eqref{kappacmin} and \eqref{eff_p2c},  we obtain the formula for the pump to comb efficiency dependence on pump amplitude normalized to minimal threshold $f_0$, normalized coupling rate $\hat\kappa_c$ and target pump amplitude $f$:
\begin{align}
\label{eff_p2c_f1}
\eta_{\rm p2c}=\frac{8\hat\kappa_cf^2}{27f_0^2}\left(f\sqrt{6\hat d_2\left(\frac{f_0^2\hat\kappa_c}{4f^2}\right)^{1/3}}-\hat d_2\right).
\end{align}
Here $\hat\kappa_c=\kappa_c/\kappa_0$, $\hat d_2=D_2/\kappa_0$ and we also used $\zeta=\frac{\pi^2 f^2}{8}$ as well.
As we fix the detuning and combine \eqref{kappacmin} and \eqref{eff_p2c_f1}, this dependence can be plotted as a colormap, having only two free parameters: $\hat d_2=D_2/\kappa_0$ and $\hat\kappa_c=\kappa_c/\kappa_0$ [see Fig. \ref{fig:eff(k)1.5}]. We can see that for fixed dispersion value there is an optimum over the coupling value [light (green) solid line in Fig. \ref{fig:eff(k)1.5}]. It can be shown that for small GVD, it tends to $\kappa_c/\kappa_0\approx4$. In the left panel of Fig. \ref{fig:eff(k)1.5} we use $f=1$ as the most illustrative case though it is not quite correct. For greater $f$ the global maximum will shift up and eventually above $d_2=1$ line. 
Right panel of figure \ref{fig:eff(k)1.5} shows the efficiency map for $f=2$. Comparing it with the left panel we can see that the maximum goes behind the $d_2=1$ with growing $f$. The other important observation is that the number of lines grows with $f$.

Another essential way to represent the efficiency map is to recalculate the $d_2$ into the number of comb lines above 3 dB level for given $f_0$. So, we present the map of pump-to-comb efficiency in coordinates $f_0$-$N_{\rm lines}$ in Fig. \ref{fig:eff(N)}. 
One can see that the pump-to-comb efficiency is reduced with the line number increase and its growth saturates with the power increase.

For some essential applications, the absolute power of the generated comb is a more critical parameter than the efficiency. The formula \eqref{Pouttotal} suggests that it does not directly depend on the pump power and increases mainly with the threshold power. The increased pump power (for fixed threshold) provides greater possible detuning values; however, this effect is weaker.

\begin{figure}
\centering
\includegraphics[width=1\linewidth]{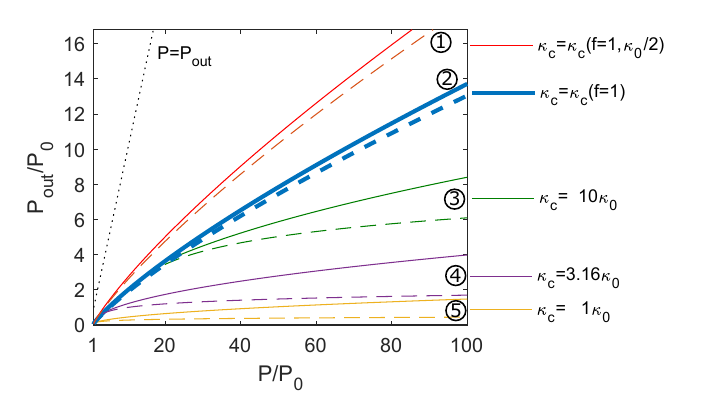}
\caption{Total comb power without pumped mode, normalized to $P_0$, vs. the normalized pump power for different coupling rates. Solid lines correspond to maximum detuning and dashed -- to locked detuning. Thick lines 1 and 2 corresponds to the threshold pumping case $f=1$ according to \eqref{kappacmin}. Lines 3, 4, 5 correspond to different fixed coupling. Line 2 corresponds to the case $f=1$ and 2 times less internal losses. The black dotted line shows $P=P_{\rm out}$, e.g. 100\% efficiency limit. System parameters $\kappa_0/2\pi=188$ MHz, $D_2/2\pi=14.3$ MHz. All quantities are plotted in dimensionless units.}
\label{fig:pout}
\end{figure}

In Fig. \ref{fig:pout}, the total comb power without pumped mode normalized to $P_0$  ($P_{\rm out}=\eta_{\rm p2c}P$) is shown depending on the normalized pump power $f_0^2$ for different coupling rates. The thick line 2 corresponds to the threshold pumping case $f=1$, so that the coupling is changed with the pump power, according to \eqref{kappacmin} (solid line for maximum detuning and dashed -- for locked detuning). We can see that this regime is always superior to the case of fixed coupling [see lines 3, 4, 5]. The lines in Fig. \ref{fig:pout} do not change if the minimal threshold power $P_0$ is changed by means of the material parameters.
If the $P_0$ is increased by means of the microresonator intrinsic loss $\kappa_0$, then the optimal curve goes up (see line 1 in fig \ref{fig:pout}). This means that though we should increase both pump and the threshold power for higher output comb power, the microresonator intrinsic losses should be never increased.

\begin{figure*}[htb]
\begin{center}
\includegraphics[width=0.49\linewidth]{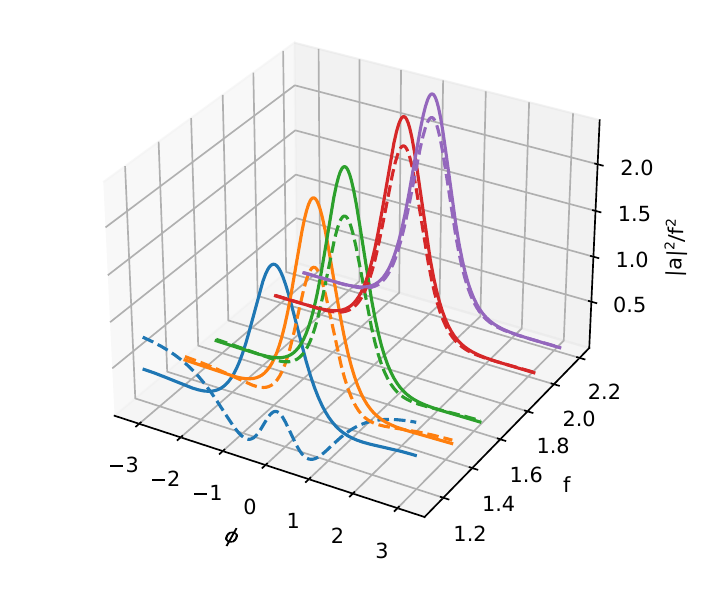}
\includegraphics[width=0.49\linewidth]{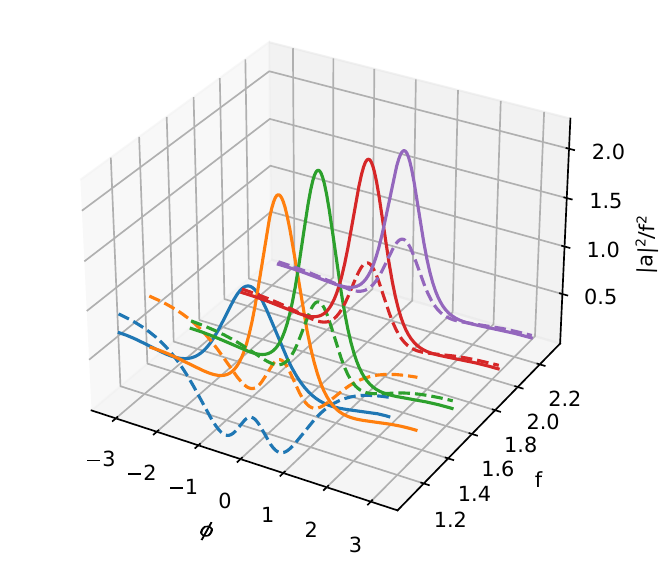}
\end{center}
\caption{Simulated soliton profiles (solid lines) and their analytical approximations (dashed lines) for the different values of the normalized pump amplitude $f$. 
Left panel shows the results for the near cut-off ($\approx 0.95\zeta^{\rm max}_{\rm eff}$) detunings and optimal dispersion coefficient \eqref{eta_maxd2}.
Right panel shows the results for the SIL detuning and optimal dispersion coefficient \eqref{eta_maxd2}. All quantities are plotted in dimensionless units.}
\label{solitons_3D}
\end{figure*}

It is also useful to add the line corresponding to a variation of the intrinsic losses to the previous efficiency maps in right panel of Figs. \ref{fig:eff(k)1.5} and \ref{fig:eff(N)}. We should note that both $N_{\rm lines}$ and $P/P_{0}$ depend on the microresonator intrinsic loss rate $\kappa_0$, and if we tune it having other parameters fixed the working point will follow the curve $\hat d_2=f_0D_2/\kappa_{\rm max}$, where $f_0\propto 1/\kappa_0$. This can be also written in terms of the comb width (in units of $D_1$)
\begin{align}
\label{Nk}
N_{\rm lines}(f_0)=\frac{4\arccosh\sqrt{2}}{\pi}\sqrt{\frac{\zeta}{D_2}\sqrt{\frac{27P\kappa_0^2}{4P_0}}}\left(1-\frac{1}{27f_0}\right).
\end{align}
This path is shown in Figs. \ref{fig:eff(N)} and \ref{fig:eff(k)1.5} for $D_2/2\pi=14.3$ MHz, $\kappa_{\rm max}/2\pi=1$ GHz with thin (pink) solid curve and an arrow shows the way of $\kappa_0$ increase. This line also represents the same configurations as the upper branch of the black line in the Fig. \ref{fig:eff(Q)}. Such analysis also shows that the high intrinsic losses are not optimal for the conversion efficiency.

{To conclude this section we have to note that though such analytical derivations may be inaccurate in the regions of small pump and detuning, they bring general relations and deeper insight to the conversion efficiency problem. Enhancements to the theory can be done in terms of including higher dispersion orders and soliton form corrections, but this can make it less transparent or analytically irresolvable making such complications not fully justified. Thus, having the regions of interest highlighted by a simple approximations, we feel reasonable to proceed directly to numerical study.}

\begin{figure*}[ht!]
\begin{center}
\includegraphics[width=0.49\linewidth]{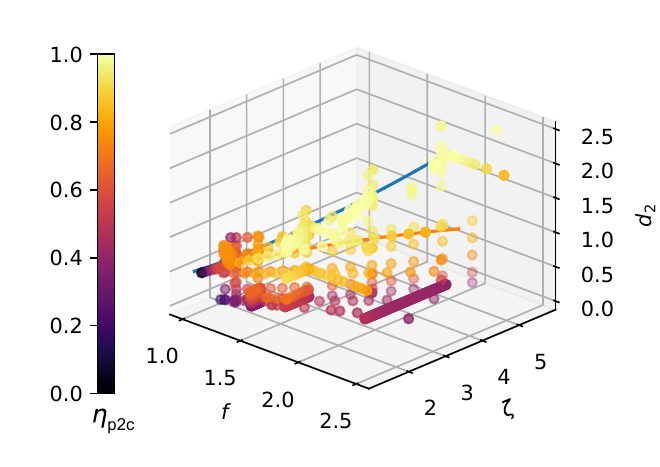}
\includegraphics[width=0.49\linewidth]{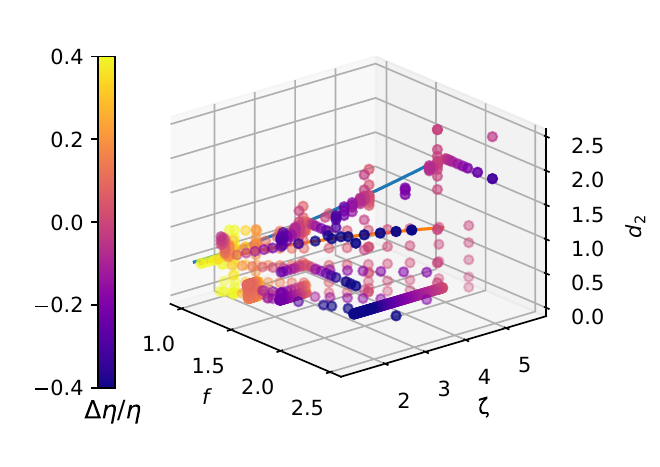}
\end{center}
\caption{Simulated pump to comb conversion efficiency (left) and its difference from the analytical approximation \eqref{eff_p2c}. The blue and orange lines correspond to the optimal dispersion value \eqref{eta_maxd2} for the near cut-off detuning and for the locked detuning. All quantities are plotted in dimensionless units.}
\label{eff_3D}
\end{figure*}

\section{Numerical modelling}
As noted in the previous sections, the analytical approximation \eqref{comblines} may be inaccurate for small detunings $\zeta$ and low pump amplitude $f$.
Here we perform the numerical modelling using the pulse propagation in Lugiato-Lefever equation formalism \cite{Chembo2013,Godey2014} and compare with the analytical solution. We also calculate numerically the efficiency to compare with \eqref{eff_p2c} for different combinations of parameters including the optimum \eqref{eta_maxd2} for cut-off and SIL detunings. In this study we find that the solution \eqref{comblines} actually diverges with the numerically obtained waveforms near cut-off detuning at low pump power.

\begin{figure*}[ht!]
\begin{center}
\includegraphics[width=0.49\linewidth]{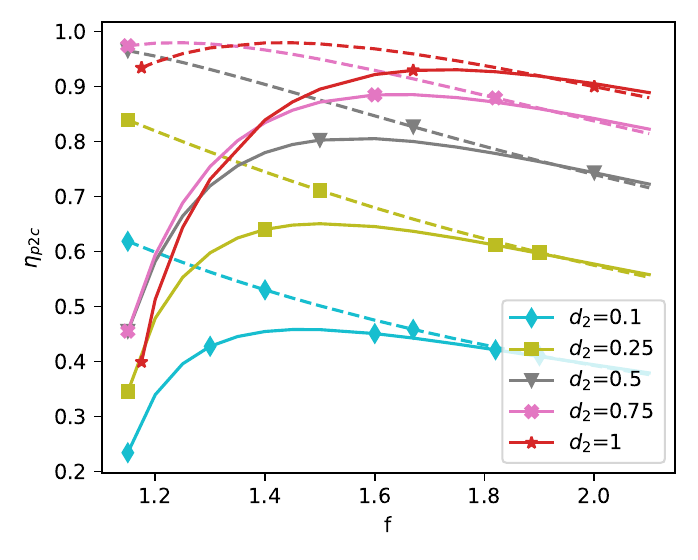}
\includegraphics[width=0.49\linewidth]{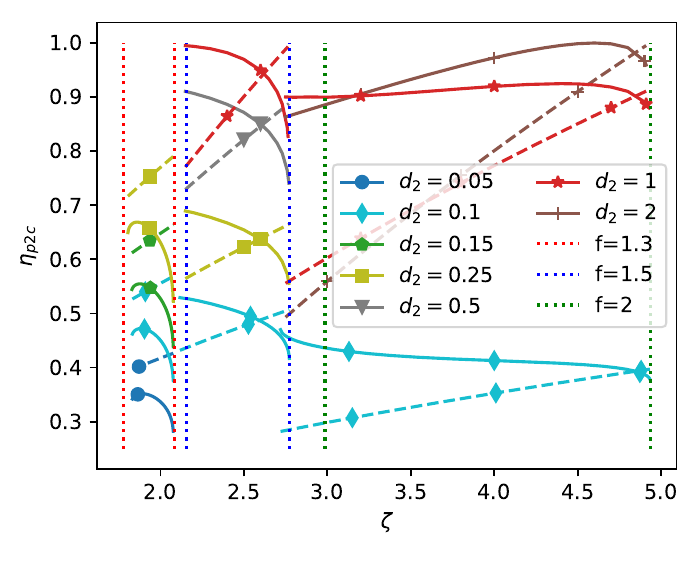}
\end{center}
\caption{Simulated pump to comb conversion efficiency (solid lines) and the analytical approximation \eqref{eff_p2c} (dashed lines) for fixed dispersion. 
Left panel shows the pump sweep for near cut-off ($\approx 0.98\zeta^{\rm max}_{\rm eff}$) detunings and different dispersion coefficients $d_2$.
Right panel shows the detuning sweep for different normalized pump amplitudes $f$ (lines inside corresponding detuning boundaries) and different dispersion coefficients $d_2$ (line colors), the dotted vertical lines show the SIL (left) and cut-off (right) detuning for the corresponding pump amplitude. Note, that while the cut-off detuning is strict for soliton existence, the SIL detuning is generally higher than lower existence boundary. All quantities are plotted in dimensionless units.}
\label{eff_2D}
\end{figure*}

For numerical analysis, we solve Lugiato-Lefever equation
\begin{equation}
\label{LLE}
     \frac{\partial a}{\partial\tau}=i\frac{d_2}{2}\frac{\partial^2 a}{\partial\varphi^2}-[1+i\zeta]a+i|a|^2a+f
\end{equation}
with hyperbolic secant initial condition and propagate until the solution reach stationary regime. Here $a$ is the intracavity field distribution over the azimuthal angle $\varphi$ (Fourier transform of the modal fields $a_\mu$) and time is normalised to the total decay rate $\tau=\kappa t/2$.
Fig. \ref{solitons_3D} shows the simulated and analytical soliton profiles for the cut-off and SIL detunings with corresponding optimal dispersion \eqref{eta_maxd2}. We can see that approaching the $f=1$ the form of the soliton does not change as significantly as the theory predicts. This suggests that the analytical soliton profile has wrong phase dependence for low $f$. For low detuning values satisfactory correspondence is not found.

Then we take the Fourier transform of $a$ and sum up all the modal powers excluding the central one. Assuming overcoupling ($\eta=1$) we are left with only three free parameters -- the detuning $\zeta$, the dispersion $d_2$ and the pump amplitude $f$, so we can build a cloud of efficiency values to be compared with \eqref{eff_p2c}.

In Fig. \ref{eff_3D} the simulated efficiency values (left) and their difference from the analytical values \eqref{eff_p2c} (right) are shown. The optimal parameter combinations \eqref{eta_maxd2} for cut-off and SIL detunings are shown with blue and orange lines.
We can see that as the pump amplitude reduces to 1 the analytic solution starts to overestimate the numerical one [see also left panel of Fig. \ref{eff_2D}]. At the same time, the values at small detunings [e.g. in the self-injection locking regime] are underestimated [see also right panel of Fig. \ref{eff_2D}]. For $f<1.5$ the efficiency is always less than 90\%. The maximal value of 100\% appears at $f=1.6$ at the SIL detuning (and optimal $d_2\approx1$) and then travels to the cut-off, reaching it near $f=2$ (and optimal $d_2=2$), where the analytical hyperbolic secant squared solution becomes precise. 
Generally, in the region of bad analytical approximation $f<2$ the optimal detunings will be slightly lower than analytically predicted and dispersion --  slightly higher. 
The highest concentration of the high-efficiency regime points is near $f\approx1.6$. 
The right panel of Fig. \ref{eff_2D} also shows the fact that the soliton existence region over detuning is different for different pump amplitudes. Note, that while the cut-off detuning is strict for soliton existence, the SIL detuning is generally higher than lower existence boundary.

{In literature one may find the next order approximation for the soliton form \cite{Guo2017,Li:18}. We should note, however, that though giving better results for lower detunings, this modification still does not work for low normalised pump amplitudes, which are of most interest, while at high detunings the simple approximation is precise.}

\section{Backward wave}
\begin{figure}[ht!]
\begin{center}
\includegraphics[width=0.9\linewidth]{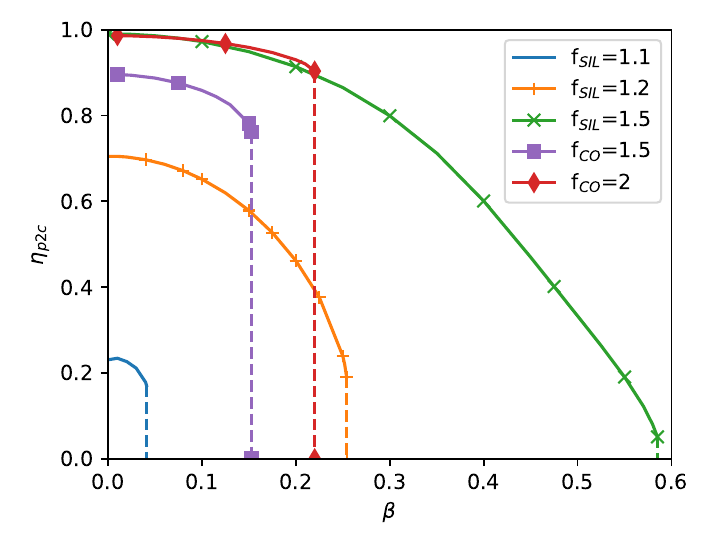}
\end{center}
\caption{Pump to comb dependence on normalised backscattering coefficient $\beta$. SIL and CO stands for SIL detuning and cut-off detuning. Dispersion is chosen optimal \eqref{eta_maxd2}. Vertical dashed lines show the backscattering-related cut-off. All quantities are plotted in dimensionless units.}
\label{eff_beta}
\end{figure}

The reader should note that though considering the SIL detunings, no simulation including the locking mechanism has actually been performed, i.e. no laser and pump frequency dynamics have been considered in \eqref{LLE}. Till now we also restrict ourselves to only forward wave. It was shown, however, that the existence of a backward wave influences the soliton generation and should be taken into consideration \cite{Fujii:17,Yang:2017,Lobanov:20,Fan:20}. This influence increases with increasing the backscattering and also with decreasing the pump power \cite{Lobanov:20}. Obviously, backscattering will reduce the efficiency, letting the power into the backward wave. Additional analysis was performed using bidirectional approach \cite{Kondratiev:20, Skryabin:20, Lobanov:20}
 \begin{align}
\label{ModelledEqs}
&\dfrac{\partial a}{\partial \tau}=i\dfrac{d_2}{2}\dfrac{\partial^2a}{\partial\varphi^2} -(1+i\zeta)a+i\left(|a|^2+2U_b\right)a+i\beta b(-\varphi)+f,\nonumber\\ 
&\dfrac{\partial b}{\partial \tau}=i\dfrac{d_2}{2}\dfrac{\partial^2b}{\partial\varphi^2}-(1+i\zeta)b+i\left(|b|^2+2U_a\right)b +i\beta a(-\varphi)\nonumber
\end{align}
for several key points of the parameter space. Here $b$ stands for the backward wave amplitude, $\beta$ is the normalised backscattering coefficient [equal to the ratio of the linear mode-splitting value and loaded microresonator linewidth], $U_a=\int |a(\varphi)|^2\frac{d\varphi}{2\pi}$ and  $U_b=\int |b(\varphi)|^2\frac{d\varphi}{2\pi}$ are average intracavity powers of the waves. 
It was revealed that for most interesting values of the pump power ($f>1.4$) the backscattering of $\beta<0.1$ do not significantly harm the efficiency. Again we see that for higher backscattering the soliton generation is hindered [see vertical dashed lines in Fig \ref{eff_beta}]. We should note that value of $\beta\gg1$ can be controlled effectively in specially designed photonic crystal ring resonators \cite{Ulanov:22}. It is also stated in \cite{Ulanov:22} that possible large values of $\beta\gg1$ allow for wider change of the effective detuning inside the locked state, which can partially compensate the loss of efficiency due to the backward wave.

\section{CONCLUSION}
In conclusion, an analytical guideline has been proposed to optimize the soliton microcomb generation efficiency in a high-Q optical microresonator. {The developed theory analyses the combination of pump amplitude and detuning, microresonator second-order dispersion and laser-to-microresonator coupling rate, and  shows theoretical solution to reaching 100\% in a simple single-resonator scheme}. The analytical calculation can predict the optimal parameters for pumps above $f=2$. For a lower pump, optimal detunings will be slightly lower and dispersion slightly higher {due to approximation failure and direct numerical calculation is encouraged}.
Generally, the system should be pumped not far from the generation threshold ($f\approx1.6$). This regime has a smaller optimal dispersion value and maximal efficiency for most detuning values and ensures no wasted pump power -- more pump power does not change the soliton amplitude and average power.
For this purpose, the coupling should be increased or the pump power should be reduced. 
In practical terms, a trade-off between the efficiency and comb line number should be chosen at the design stage, forcing the appropriate dispersion profile to be tailored (by means of material and geometry engineering) for an achievable coupling. However, it should be noted that dispersion engineering can also affect the threshold power and coupling due to the changes in the effective mode volume and effective index. 
Additionally, the regime of the self-injection locking of the pump laser can help to maintain the necessary pump frequency detuning since, at threshold pumping, the soliton existence range in the spectral domain is quite narrow. Though the analytical theory was found to fail in the low pump -- low detuning region, numerical simulations show the optimum point near $f\approx1.6$, $d_2\approx0.95$.
Finally, to increase the absolute comb power, the same strategy should be applied, with the only difference being that the generation threshold power should be maximized. To this aim, the safest approach is to increase the coupling efficiency significantly while keeping the intrinsic losses low.
\\
\section*{Acknowledgments}
RQC team thanks Russian Science Foundation (Project No. 23-42-00111). V.E.L. acknowledges personal support from the Foundation for the Advancement of Theoretical Physics and Mathematics “BASIS”.

\section*{Competing interests}
The authors declare no competing interests.


\bibliography{main}

\begin{thebibliography}{55}%
\makeatletter
\providecommand \@ifxundefined [1]{%
 \@ifx{#1\undefined}
}%
\providecommand \@ifnum [1]{%
 \ifnum #1\expandafter \@firstoftwo
 \else \expandafter \@secondoftwo
 \fi
}%
\providecommand \@ifx [1]{%
 \ifx #1\expandafter \@firstoftwo
 \else \expandafter \@secondoftwo
 \fi
}%
\providecommand \natexlab [1]{#1}%
\providecommand \enquote  [1]{``#1''}%
\providecommand \bibnamefont  [1]{#1}%
\providecommand \bibfnamefont [1]{#1}%
\providecommand \citenamefont [1]{#1}%
\providecommand \href@noop [0]{\@secondoftwo}%
\providecommand \href [0]{\begingroup \@sanitize@url \@href}%
\providecommand \@href[1]{\@@startlink{#1}\@@href}%
\providecommand \@@href[1]{\endgroup#1\@@endlink}%
\providecommand \@sanitize@url [0]{\catcode `\\12\catcode `\$12\catcode
  `\&12\catcode `\#12\catcode `\^12\catcode `\_12\catcode `\%12\relax}%
\providecommand \@@startlink[1]{}%
\providecommand \@@endlink[0]{}%
\providecommand \url  [0]{\begingroup\@sanitize@url \@url }%
\providecommand \@url [1]{\endgroup\@href {#1}{\urlprefix }}%
\providecommand \urlprefix  [0]{URL }%
\providecommand \Eprint [0]{\href }%
\providecommand \doibase [0]{http://dx.doi.org/}%
\providecommand \selectlanguage [0]{\@gobble}%
\providecommand \bibinfo  [0]{\@secondoftwo}%
\providecommand \bibfield  [0]{\@secondoftwo}%
\providecommand \translation [1]{[#1]}%
\providecommand \BibitemOpen [0]{}%
\providecommand \bibitemStop [0]{}%
\providecommand \bibitemNoStop [0]{.\EOS\space}%
\providecommand \EOS [0]{\spacefactor3000\relax}%
\providecommand \BibitemShut  [1]{\csname bibitem#1\endcsname}%
\let\auto@bib@innerbib\@empty
\bibitem [{\citenamefont {Kippenberg}\ \emph {et~al.}(2011)\citenamefont
  {Kippenberg}, \citenamefont {Holzwarth},\ and\ \citenamefont
  {Diddams}}]{Kippenberg2011}%
  \BibitemOpen
  \bibfield  {author} {\bibinfo {author} {\bibfnamefont {T.~J.}\ \bibnamefont
  {Kippenberg}}, \bibinfo {author} {\bibfnamefont {R.}~\bibnamefont
  {Holzwarth}}, \ and\ \bibinfo {author} {\bibfnamefont {S.~A.}\ \bibnamefont
  {Diddams}},\ }\href {\doibase 10.1126/science.1193968} {\bibfield  {journal}
  {\bibinfo  {journal} {Science}\ }\textbf {\bibinfo {volume} {332}},\ \bibinfo
  {pages} {555} (\bibinfo {year} {2011})}\BibitemShut {NoStop}%
\bibitem [{\citenamefont {Chembo}(2016)}]{Chembo:2016}%
  \BibitemOpen
  \bibfield  {author} {\bibinfo {author} {\bibfnamefont {Y.~K.}\ \bibnamefont
  {Chembo}},\ }\href {\doibase doi:10.1515/nanoph-2016-0013} {\bibfield
  {journal} {\bibinfo  {journal} {Nanophotonics}\ }\textbf {\bibinfo {volume}
  {5}},\ \bibinfo {pages} {214} (\bibinfo {year} {2016})}\BibitemShut {NoStop}%
\bibitem [{\citenamefont {Pasquazi}\ \emph {et~al.}(2018)\citenamefont
  {Pasquazi}, \citenamefont {Peccianti}, \citenamefont {Razzari}, \citenamefont
  {Moss}, \citenamefont {Coen}, \citenamefont {Erkintalo}, \citenamefont
  {Chembo}, \citenamefont {Hansson}, \citenamefont {Wabnitz}, \citenamefont
  {Del’Haye}, \citenamefont {Xue}, \citenamefont {Weiner},\ and\
  \citenamefont {Morandotti}}]{PASQUAZI20181}%
  \BibitemOpen
  \bibfield  {author} {\bibinfo {author} {\bibfnamefont {A.}~\bibnamefont
  {Pasquazi}}, \bibinfo {author} {\bibfnamefont {M.}~\bibnamefont {Peccianti}},
  \bibinfo {author} {\bibfnamefont {L.}~\bibnamefont {Razzari}}, \bibinfo
  {author} {\bibfnamefont {D.~J.}\ \bibnamefont {Moss}}, \bibinfo {author}
  {\bibfnamefont {S.}~\bibnamefont {Coen}}, \bibinfo {author} {\bibfnamefont
  {M.}~\bibnamefont {Erkintalo}}, \bibinfo {author} {\bibfnamefont {Y.~K.}\
  \bibnamefont {Chembo}}, \bibinfo {author} {\bibfnamefont {T.}~\bibnamefont
  {Hansson}}, \bibinfo {author} {\bibfnamefont {S.}~\bibnamefont {Wabnitz}},
  \bibinfo {author} {\bibfnamefont {P.}~\bibnamefont {Del’Haye}}, \bibinfo
  {author} {\bibfnamefont {X.}~\bibnamefont {Xue}}, \bibinfo {author}
  {\bibfnamefont {A.~M.}\ \bibnamefont {Weiner}}, \ and\ \bibinfo {author}
  {\bibfnamefont {R.}~\bibnamefont {Morandotti}},\ }\href {\doibase
  https://doi.org/10.1016/j.physrep.2017.08.004} {\bibfield  {journal}
  {\bibinfo  {journal} {Physics Reports}\ }\textbf {\bibinfo {volume} {729}},\
  \bibinfo {pages} {1} (\bibinfo {year} {2018})},\ \bibinfo {note}
  {micro-combs: A novel generation of optical sources}\BibitemShut {NoStop}%
\bibitem [{\citenamefont {Gaeta}\ \emph {et~al.}(2019)\citenamefont {Gaeta},
  \citenamefont {Lipson},\ and\ \citenamefont {Kippenberg}}]{Gaeta2019}%
  \BibitemOpen
  \bibfield  {author} {\bibinfo {author} {\bibfnamefont {A.~L.}\ \bibnamefont
  {Gaeta}}, \bibinfo {author} {\bibfnamefont {M.}~\bibnamefont {Lipson}}, \
  and\ \bibinfo {author} {\bibfnamefont {T.~J.}\ \bibnamefont {Kippenberg}},\
  }\href {\doibase 10.1038/s41566-019-0358-x} {\bibfield  {journal} {\bibinfo
  {journal} {Nature Photonics}\ }\textbf {\bibinfo {volume} {13}},\ \bibinfo
  {pages} {158} (\bibinfo {year} {2019})}\BibitemShut {NoStop}%
\bibitem [{\citenamefont {Kovach}\ \emph {et~al.}(2020)\citenamefont {Kovach},
  \citenamefont {Chen}, \citenamefont {He}, \citenamefont {Choi}, \citenamefont
  {Dogan}, \citenamefont {Ghasemkhani}, \citenamefont {Taheri},\ and\
  \citenamefont {Armani}}]{Kovach:20}%
  \BibitemOpen
  \bibfield  {author} {\bibinfo {author} {\bibfnamefont {A.}~\bibnamefont
  {Kovach}}, \bibinfo {author} {\bibfnamefont {D.}~\bibnamefont {Chen}},
  \bibinfo {author} {\bibfnamefont {J.}~\bibnamefont {He}}, \bibinfo {author}
  {\bibfnamefont {H.}~\bibnamefont {Choi}}, \bibinfo {author} {\bibfnamefont
  {A.~H.}\ \bibnamefont {Dogan}}, \bibinfo {author} {\bibfnamefont
  {M.}~\bibnamefont {Ghasemkhani}}, \bibinfo {author} {\bibfnamefont
  {H.}~\bibnamefont {Taheri}}, \ and\ \bibinfo {author} {\bibfnamefont {A.~M.}\
  \bibnamefont {Armani}},\ }\href {\doibase 10.1364/AOP.376924} {\bibfield
  {journal} {\bibinfo  {journal} {Adv. Opt. Photon.}\ }\textbf {\bibinfo
  {volume} {12}},\ \bibinfo {pages} {135} (\bibinfo {year} {2020})}\BibitemShut
  {NoStop}%
\bibitem [{\citenamefont {Nie}\ \emph {et~al.}(2022)\citenamefont {Nie},
  \citenamefont {Xie}, \citenamefont {Li},\ and\ \citenamefont
  {Huang}}]{Nie:2022}%
  \BibitemOpen
  \bibfield  {author} {\bibinfo {author} {\bibfnamefont {M.}~\bibnamefont
  {Nie}}, \bibinfo {author} {\bibfnamefont {Y.}~\bibnamefont {Xie}}, \bibinfo
  {author} {\bibfnamefont {B.}~\bibnamefont {Li}}, \ and\ \bibinfo {author}
  {\bibfnamefont {S.-W.}\ \bibnamefont {Huang}},\ }\href {\doibase
  https://doi.org/10.1016/j.pquantelec.2022.100437} {\bibfield  {journal}
  {\bibinfo  {journal} {Progress in Quantum Electronics}\ }\textbf {\bibinfo
  {volume} {86}},\ \bibinfo {pages} {100437} (\bibinfo {year}
  {2022})}\BibitemShut {NoStop}%
\bibitem [{\citenamefont {Suh}\ \emph {et~al.}(2016)\citenamefont {Suh},
  \citenamefont {Yang}, \citenamefont {Yang}, \citenamefont {Yi},\ and\
  \citenamefont {Vahala}}]{Suh600}%
  \BibitemOpen
  \bibfield  {author} {\bibinfo {author} {\bibfnamefont {M.-G.}\ \bibnamefont
  {Suh}}, \bibinfo {author} {\bibfnamefont {Q.-F.}\ \bibnamefont {Yang}},
  \bibinfo {author} {\bibfnamefont {K.~Y.}\ \bibnamefont {Yang}}, \bibinfo
  {author} {\bibfnamefont {X.}~\bibnamefont {Yi}}, \ and\ \bibinfo {author}
  {\bibfnamefont {K.~J.}\ \bibnamefont {Vahala}},\ }\href {\doibase
  10.1126/science.aah6516} {\bibfield  {journal} {\bibinfo  {journal}
  {Science}\ }\textbf {\bibinfo {volume} {354}},\ \bibinfo {pages} {600}
  (\bibinfo {year} {2016})}\BibitemShut {NoStop}%
\bibitem [{\citenamefont {Kues}\ \emph {et~al.}(2019)\citenamefont {Kues},
  \citenamefont {Reimer}, \citenamefont {Lukens}, \citenamefont {Munro},
  \citenamefont {Weiner}, \citenamefont {Moss},\ and\ \citenamefont
  {Morandotti}}]{Kues2019}%
  \BibitemOpen
  \bibfield  {author} {\bibinfo {author} {\bibfnamefont {M.}~\bibnamefont
  {Kues}}, \bibinfo {author} {\bibfnamefont {C.}~\bibnamefont {Reimer}},
  \bibinfo {author} {\bibfnamefont {J.~M.}\ \bibnamefont {Lukens}}, \bibinfo
  {author} {\bibfnamefont {W.~J.}\ \bibnamefont {Munro}}, \bibinfo {author}
  {\bibfnamefont {A.~M.}\ \bibnamefont {Weiner}}, \bibinfo {author}
  {\bibfnamefont {D.~J.}\ \bibnamefont {Moss}}, \ and\ \bibinfo {author}
  {\bibfnamefont {R.}~\bibnamefont {Morandotti}},\ }\href {\doibase
  10.1038/s41566-019-0363-0} {\bibfield  {journal} {\bibinfo  {journal} {Nature
  Photonics}\ }\textbf {\bibinfo {volume} {13}},\ \bibinfo {pages} {170}
  (\bibinfo {year} {2019})}\BibitemShut {NoStop}%
\bibitem [{\citenamefont {Riemensberger}\ \emph {et~al.}(2020)\citenamefont
  {Riemensberger}, \citenamefont {Lukashchuk}, \citenamefont {Karpov},
  \citenamefont {Weng}, \citenamefont {Lucas}, \citenamefont {Liu},\ and\
  \citenamefont {Kippenberg}}]{Riemensberger2020}%
  \BibitemOpen
  \bibfield  {author} {\bibinfo {author} {\bibfnamefont {J.}~\bibnamefont
  {Riemensberger}}, \bibinfo {author} {\bibfnamefont {A.}~\bibnamefont
  {Lukashchuk}}, \bibinfo {author} {\bibfnamefont {M.}~\bibnamefont {Karpov}},
  \bibinfo {author} {\bibfnamefont {W.}~\bibnamefont {Weng}}, \bibinfo {author}
  {\bibfnamefont {E.}~\bibnamefont {Lucas}}, \bibinfo {author} {\bibfnamefont
  {J.}~\bibnamefont {Liu}}, \ and\ \bibinfo {author} {\bibfnamefont {T.~J.}\
  \bibnamefont {Kippenberg}},\ }\href {\doibase 10.1038/s41586-020-2239-3}
  {\bibfield  {journal} {\bibinfo  {journal} {Nature}\ }\textbf {\bibinfo
  {volume} {581}},\ \bibinfo {pages} {164} (\bibinfo {year}
  {2020})}\BibitemShut {NoStop}%
\bibitem [{\citenamefont {Marin-Palomo}\ \emph {et~al.}(2017)\citenamefont
  {Marin-Palomo}, \citenamefont {Kemal}, \citenamefont {Karpov}, \citenamefont
  {Kordts}, \citenamefont {Pfeifle}, \citenamefont {Pfeiffer}, \citenamefont
  {Trocha}, \citenamefont {Wolf}, \citenamefont {Brasch}, \citenamefont
  {Anderson}, \citenamefont {Rosenberger}, \citenamefont {Vijayan},
  \citenamefont {Freude}, \citenamefont {Kippenberg},\ and\ \citenamefont
  {Koos}}]{Marin-Palomo2017}%
  \BibitemOpen
  \bibfield  {author} {\bibinfo {author} {\bibfnamefont {P.}~\bibnamefont
  {Marin-Palomo}}, \bibinfo {author} {\bibfnamefont {J.~N.}\ \bibnamefont
  {Kemal}}, \bibinfo {author} {\bibfnamefont {M.}~\bibnamefont {Karpov}},
  \bibinfo {author} {\bibfnamefont {A.}~\bibnamefont {Kordts}}, \bibinfo
  {author} {\bibfnamefont {J.}~\bibnamefont {Pfeifle}}, \bibinfo {author}
  {\bibfnamefont {M.~H.~P.}\ \bibnamefont {Pfeiffer}}, \bibinfo {author}
  {\bibfnamefont {P.}~\bibnamefont {Trocha}}, \bibinfo {author} {\bibfnamefont
  {S.}~\bibnamefont {Wolf}}, \bibinfo {author} {\bibfnamefont {V.}~\bibnamefont
  {Brasch}}, \bibinfo {author} {\bibfnamefont {M.~H.}\ \bibnamefont
  {Anderson}}, \bibinfo {author} {\bibfnamefont {R.}~\bibnamefont
  {Rosenberger}}, \bibinfo {author} {\bibfnamefont {K.}~\bibnamefont
  {Vijayan}}, \bibinfo {author} {\bibfnamefont {W.}~\bibnamefont {Freude}},
  \bibinfo {author} {\bibfnamefont {T.~J.}\ \bibnamefont {Kippenberg}}, \ and\
  \bibinfo {author} {\bibfnamefont {C.}~\bibnamefont {Koos}},\ }\href {\doibase
  10.1038/nature22387} {\bibfield  {journal} {\bibinfo  {journal} {Nature}\
  }\textbf {\bibinfo {volume} {546}},\ \bibinfo {pages} {274} (\bibinfo {year}
  {2017})}\BibitemShut {NoStop}%
\bibitem [{\citenamefont {Suh}\ \emph {et~al.}(2019)\citenamefont {Suh},
  \citenamefont {Yi}, \citenamefont {Lai}, \citenamefont {Leifer},
  \citenamefont {Grudinin}, \citenamefont {Vasisht}, \citenamefont {Martin},
  \citenamefont {Fitzgerald}, \citenamefont {Doppmann}, \citenamefont {Wang},
  \citenamefont {Mawet}, \citenamefont {Papp}, \citenamefont {Diddams},
  \citenamefont {Beichman},\ and\ \citenamefont {Vahala}}]{Suh2019}%
  \BibitemOpen
  \bibfield  {author} {\bibinfo {author} {\bibfnamefont {M.-G.}\ \bibnamefont
  {Suh}}, \bibinfo {author} {\bibfnamefont {X.}~\bibnamefont {Yi}}, \bibinfo
  {author} {\bibfnamefont {Y.-H.}\ \bibnamefont {Lai}}, \bibinfo {author}
  {\bibfnamefont {S.}~\bibnamefont {Leifer}}, \bibinfo {author} {\bibfnamefont
  {I.~S.}\ \bibnamefont {Grudinin}}, \bibinfo {author} {\bibfnamefont
  {G.}~\bibnamefont {Vasisht}}, \bibinfo {author} {\bibfnamefont {E.~C.}\
  \bibnamefont {Martin}}, \bibinfo {author} {\bibfnamefont {M.~P.}\
  \bibnamefont {Fitzgerald}}, \bibinfo {author} {\bibfnamefont
  {G.}~\bibnamefont {Doppmann}}, \bibinfo {author} {\bibfnamefont
  {J.}~\bibnamefont {Wang}}, \bibinfo {author} {\bibfnamefont {D.}~\bibnamefont
  {Mawet}}, \bibinfo {author} {\bibfnamefont {S.~B.}\ \bibnamefont {Papp}},
  \bibinfo {author} {\bibfnamefont {S.~A.}\ \bibnamefont {Diddams}}, \bibinfo
  {author} {\bibfnamefont {C.}~\bibnamefont {Beichman}}, \ and\ \bibinfo
  {author} {\bibfnamefont {K.}~\bibnamefont {Vahala}},\ }\href {\doibase
  10.1038/s41566-018-0312-3} {\bibfield  {journal} {\bibinfo  {journal} {Nature
  Photonics}\ }\textbf {\bibinfo {volume} {13}},\ \bibinfo {pages} {25}
  (\bibinfo {year} {2019})}\BibitemShut {NoStop}%
\bibitem [{\citenamefont {Kippenberg}\ \emph {et~al.}(2018)\citenamefont
  {Kippenberg}, \citenamefont {Gaeta}, \citenamefont {Lipson},\ and\
  \citenamefont {Gorodetsky}}]{Kippenbergeaan8083}%
  \BibitemOpen
  \bibfield  {author} {\bibinfo {author} {\bibfnamefont {T.~J.}\ \bibnamefont
  {Kippenberg}}, \bibinfo {author} {\bibfnamefont {A.~L.}\ \bibnamefont
  {Gaeta}}, \bibinfo {author} {\bibfnamefont {M.}~\bibnamefont {Lipson}}, \
  and\ \bibinfo {author} {\bibfnamefont {M.~L.}\ \bibnamefont {Gorodetsky}},\
  }\href {\doibase 10.1126/science.aan8083} {\bibfield  {journal} {\bibinfo
  {journal} {Science}\ }\textbf {\bibinfo {volume} {361}} (\bibinfo {year}
  {2018}),\ 10.1126/science.aan8083}\BibitemShut {NoStop}%
\bibitem [{\citenamefont {Sun}\ \emph {et~al.}(2023)\citenamefont {Sun},
  \citenamefont {Wu}, \citenamefont {Tan}, \citenamefont {Xu}, \citenamefont
  {Li}, \citenamefont {Morandotti}, \citenamefont {Mitchell},\ and\
  \citenamefont {Moss}}]{Sun:23}%
  \BibitemOpen
  \bibfield  {author} {\bibinfo {author} {\bibfnamefont {Y.}~\bibnamefont
  {Sun}}, \bibinfo {author} {\bibfnamefont {J.}~\bibnamefont {Wu}}, \bibinfo
  {author} {\bibfnamefont {M.}~\bibnamefont {Tan}}, \bibinfo {author}
  {\bibfnamefont {X.}~\bibnamefont {Xu}}, \bibinfo {author} {\bibfnamefont
  {Y.}~\bibnamefont {Li}}, \bibinfo {author} {\bibfnamefont {R.}~\bibnamefont
  {Morandotti}}, \bibinfo {author} {\bibfnamefont {A.}~\bibnamefont
  {Mitchell}}, \ and\ \bibinfo {author} {\bibfnamefont {D.~J.}\ \bibnamefont
  {Moss}},\ }\href {\doibase 10.1364/AOP.470264} {\bibfield  {journal}
  {\bibinfo  {journal} {Adv. Opt. Photon.}\ }\textbf {\bibinfo {volume} {15}},\
  \bibinfo {pages} {86} (\bibinfo {year} {2023})}\BibitemShut {NoStop}%
\bibitem [{\citenamefont {Karpov}\ \emph {et~al.}(2019)\citenamefont {Karpov},
  \citenamefont {Pfeiffer}, \citenamefont {Guo}, \citenamefont {Weng},
  \citenamefont {Liu},\ and\ \citenamefont {Kippenberg}}]{Karpov2019}%
  \BibitemOpen
  \bibfield  {author} {\bibinfo {author} {\bibfnamefont {M.}~\bibnamefont
  {Karpov}}, \bibinfo {author} {\bibfnamefont {M.~H.~P.}\ \bibnamefont
  {Pfeiffer}}, \bibinfo {author} {\bibfnamefont {H.}~\bibnamefont {Guo}},
  \bibinfo {author} {\bibfnamefont {W.}~\bibnamefont {Weng}}, \bibinfo {author}
  {\bibfnamefont {J.}~\bibnamefont {Liu}}, \ and\ \bibinfo {author}
  {\bibfnamefont {T.~J.}\ \bibnamefont {Kippenberg}},\ }\href {\doibase
  10.1038/s41567-019-0635-0} {\bibfield  {journal} {\bibinfo  {journal} {Nature
  Physics}\ }\textbf {\bibinfo {volume} {15}},\ \bibinfo {pages} {1071}
  (\bibinfo {year} {2019})}\BibitemShut {NoStop}%
\bibitem [{\citenamefont {Xue}\ \emph {et~al.}(2019)\citenamefont {Xue},
  \citenamefont {Zheng},\ and\ \citenamefont {Zhou}}]{Xue2019}%
  \BibitemOpen
  \bibfield  {author} {\bibinfo {author} {\bibfnamefont {X.}~\bibnamefont
  {Xue}}, \bibinfo {author} {\bibfnamefont {X.}~\bibnamefont {Zheng}}, \ and\
  \bibinfo {author} {\bibfnamefont {B.}~\bibnamefont {Zhou}},\ }\href {\doibase
  10.1038/s41566-019-0436-0} {\bibfield  {journal} {\bibinfo  {journal} {Nature
  Photonics}\ }\textbf {\bibinfo {volume} {13}},\ \bibinfo {pages} {616}
  (\bibinfo {year} {2019})}\BibitemShut {NoStop}%
\bibitem [{\citenamefont {Boggio}\ \emph {et~al.}(2022)\citenamefont {Boggio},
  \citenamefont {Bodenm{\"u}ller}, \citenamefont {Ahmed}, \citenamefont
  {Wabnitz}, \citenamefont {Modotto},\ and\ \citenamefont
  {Hansson}}]{Boggio2022}%
  \BibitemOpen
  \bibfield  {author} {\bibinfo {author} {\bibfnamefont {J.~M.~C.}\
  \bibnamefont {Boggio}}, \bibinfo {author} {\bibfnamefont {D.}~\bibnamefont
  {Bodenm{\"u}ller}}, \bibinfo {author} {\bibfnamefont {S.}~\bibnamefont
  {Ahmed}}, \bibinfo {author} {\bibfnamefont {S.}~\bibnamefont {Wabnitz}},
  \bibinfo {author} {\bibfnamefont {D.}~\bibnamefont {Modotto}}, \ and\
  \bibinfo {author} {\bibfnamefont {T.}~\bibnamefont {Hansson}},\ }\href
  {\doibase 10.1038/s41467-022-28927-z} {\bibfield  {journal} {\bibinfo
  {journal} {Nature Communications}\ }\textbf {\bibinfo {volume} {13}},\
  \bibinfo {pages} {1292} (\bibinfo {year} {2022})}\BibitemShut {NoStop}%
\bibitem [{\citenamefont {Xue}\ \emph {et~al.}(2017)\citenamefont {Xue},
  \citenamefont {Wang}, \citenamefont {Xuan}, \citenamefont {Qi},\ and\
  \citenamefont {Weiner}}]{Xue:17}%
  \BibitemOpen
  \bibfield  {author} {\bibinfo {author} {\bibfnamefont {X.}~\bibnamefont
  {Xue}}, \bibinfo {author} {\bibfnamefont {P.-H.}\ \bibnamefont {Wang}},
  \bibinfo {author} {\bibfnamefont {Y.}~\bibnamefont {Xuan}}, \bibinfo {author}
  {\bibfnamefont {M.}~\bibnamefont {Qi}}, \ and\ \bibinfo {author}
  {\bibfnamefont {A.~M.}\ \bibnamefont {Weiner}},\ }\href {\doibase
  10.1002/lpor.201600276} {\bibfield  {journal} {\bibinfo  {journal} {Laser \&
  Photonics Reviews}\ }\textbf {\bibinfo {volume} {11}},\ \bibinfo {pages}
  {1600276} (\bibinfo {year} {2017})}\BibitemShut {NoStop}%
\bibitem [{\citenamefont {Kondratiev}\ \emph {et~al.}(2020)\citenamefont
  {Kondratiev}, \citenamefont {Lobanov}, \citenamefont {Lonshakov},
  \citenamefont {Dmitriev}, \citenamefont {Voloshin},\ and\ \citenamefont
  {Bilenko}}]{KondratievNum:20}%
  \BibitemOpen
  \bibfield  {author} {\bibinfo {author} {\bibfnamefont {N.~M.}\ \bibnamefont
  {Kondratiev}}, \bibinfo {author} {\bibfnamefont {V.~E.}\ \bibnamefont
  {Lobanov}}, \bibinfo {author} {\bibfnamefont {E.~A.}\ \bibnamefont
  {Lonshakov}}, \bibinfo {author} {\bibfnamefont {N.~Y.}\ \bibnamefont
  {Dmitriev}}, \bibinfo {author} {\bibfnamefont {A.~S.}\ \bibnamefont
  {Voloshin}}, \ and\ \bibinfo {author} {\bibfnamefont {I.~A.}\ \bibnamefont
  {Bilenko}},\ }\href {\doibase 10.1364/OE.411544} {\bibfield  {journal}
  {\bibinfo  {journal} {Opt. Express}\ }\textbf {\bibinfo {volume} {28}},\
  \bibinfo {pages} {38892} (\bibinfo {year} {2020})}\BibitemShut {NoStop}%
\bibitem [{\citenamefont {Kim}\ \emph {et~al.}(2019)\citenamefont {Kim},
  \citenamefont {Okawachi}, \citenamefont {Jang}, \citenamefont {Yu},
  \citenamefont {Ji}, \citenamefont {Zhao}, \citenamefont {Joshi},
  \citenamefont {Lipson},\ and\ \citenamefont {Gaeta}}]{Kim:19}%
  \BibitemOpen
  \bibfield  {author} {\bibinfo {author} {\bibfnamefont {B.~Y.}\ \bibnamefont
  {Kim}}, \bibinfo {author} {\bibfnamefont {Y.}~\bibnamefont {Okawachi}},
  \bibinfo {author} {\bibfnamefont {J.~K.}\ \bibnamefont {Jang}}, \bibinfo
  {author} {\bibfnamefont {M.}~\bibnamefont {Yu}}, \bibinfo {author}
  {\bibfnamefont {X.}~\bibnamefont {Ji}}, \bibinfo {author} {\bibfnamefont
  {Y.}~\bibnamefont {Zhao}}, \bibinfo {author} {\bibfnamefont {C.}~\bibnamefont
  {Joshi}}, \bibinfo {author} {\bibfnamefont {M.}~\bibnamefont {Lipson}}, \
  and\ \bibinfo {author} {\bibfnamefont {A.~L.}\ \bibnamefont {Gaeta}},\ }\href
  {\doibase 10.1364/OL.44.004475} {\bibfield  {journal} {\bibinfo  {journal}
  {Opt. Lett.}\ }\textbf {\bibinfo {volume} {44}},\ \bibinfo {pages} {4475}
  (\bibinfo {year} {2019})}\BibitemShut {NoStop}%
\bibitem [{\citenamefont {Ji}\ \emph {et~al.}(2023)\citenamefont {Ji},
  \citenamefont {Jin}, \citenamefont {Wu}, \citenamefont {Yu}, \citenamefont
  {Yuan}, \citenamefont {Zhang}, \citenamefont {Gao}, \citenamefont {Li},
  \citenamefont {Wang}, \citenamefont {Xiang}, \citenamefont {Guo},
  \citenamefont {Feshali}, \citenamefont {Paniccia}, \citenamefont {Ilchenko},
  \citenamefont {Matsko}, \citenamefont {Bowers},\ and\ \citenamefont
  {Vahala}}]{Ji:23}%
  \BibitemOpen
  \bibfield  {author} {\bibinfo {author} {\bibfnamefont {Q.-X.}\ \bibnamefont
  {Ji}}, \bibinfo {author} {\bibfnamefont {W.}~\bibnamefont {Jin}}, \bibinfo
  {author} {\bibfnamefont {L.}~\bibnamefont {Wu}}, \bibinfo {author}
  {\bibfnamefont {Y.}~\bibnamefont {Yu}}, \bibinfo {author} {\bibfnamefont
  {Z.}~\bibnamefont {Yuan}}, \bibinfo {author} {\bibfnamefont {W.}~\bibnamefont
  {Zhang}}, \bibinfo {author} {\bibfnamefont {M.}~\bibnamefont {Gao}}, \bibinfo
  {author} {\bibfnamefont {B.}~\bibnamefont {Li}}, \bibinfo {author}
  {\bibfnamefont {H.}~\bibnamefont {Wang}}, \bibinfo {author} {\bibfnamefont
  {C.}~\bibnamefont {Xiang}}, \bibinfo {author} {\bibfnamefont
  {J.}~\bibnamefont {Guo}}, \bibinfo {author} {\bibfnamefont {A.}~\bibnamefont
  {Feshali}}, \bibinfo {author} {\bibfnamefont {M.}~\bibnamefont {Paniccia}},
  \bibinfo {author} {\bibfnamefont {V.~S.}\ \bibnamefont {Ilchenko}}, \bibinfo
  {author} {\bibfnamefont {A.~B.}\ \bibnamefont {Matsko}}, \bibinfo {author}
  {\bibfnamefont {J.~E.}\ \bibnamefont {Bowers}}, \ and\ \bibinfo {author}
  {\bibfnamefont {K.~J.}\ \bibnamefont {Vahala}},\ }\href {\doibase
  10.1364/OPTICA.478710} {\bibfield  {journal} {\bibinfo  {journal} {Optica}\
  }\textbf {\bibinfo {volume} {10}},\ \bibinfo {pages} {279} (\bibinfo {year}
  {2023})}\BibitemShut {NoStop}%
\bibitem [{\citenamefont {Li}\ \emph {et~al.}(2022)\citenamefont {Li},
  \citenamefont {Bao}, \citenamefont {Ji}, \citenamefont {Wang}, \citenamefont
  {Wu}, \citenamefont {Leifer}, \citenamefont {Beichman},\ and\ \citenamefont
  {Vahala}}]{Li:22}%
  \BibitemOpen
  \bibfield  {author} {\bibinfo {author} {\bibfnamefont {J.}~\bibnamefont
  {Li}}, \bibinfo {author} {\bibfnamefont {C.}~\bibnamefont {Bao}}, \bibinfo
  {author} {\bibfnamefont {Q.-X.}\ \bibnamefont {Ji}}, \bibinfo {author}
  {\bibfnamefont {H.}~\bibnamefont {Wang}}, \bibinfo {author} {\bibfnamefont
  {L.}~\bibnamefont {Wu}}, \bibinfo {author} {\bibfnamefont {S.}~\bibnamefont
  {Leifer}}, \bibinfo {author} {\bibfnamefont {C.}~\bibnamefont {Beichman}}, \
  and\ \bibinfo {author} {\bibfnamefont {K.}~\bibnamefont {Vahala}},\ }\href
  {\doibase 10.1364/OPTICA.443060} {\bibfield  {journal} {\bibinfo  {journal}
  {Optica}\ }\textbf {\bibinfo {volume} {9}},\ \bibinfo {pages} {231} (\bibinfo
  {year} {2022})}\BibitemShut {NoStop}%
\bibitem [{\citenamefont {Dmitriev}\ \emph {et~al.}(2022)\citenamefont
  {Dmitriev}, \citenamefont {Koptyaev}, \citenamefont {Voloshin}, \citenamefont
  {Kondratiev}, \citenamefont {Min'kov}, \citenamefont {Lobanov}, \citenamefont
  {Ryabko}, \citenamefont {Polonsky},\ and\ \citenamefont
  {Bilenko}}]{dmitriev2021hybrid}%
  \BibitemOpen
  \bibfield  {author} {\bibinfo {author} {\bibfnamefont {N.~Y.}\ \bibnamefont
  {Dmitriev}}, \bibinfo {author} {\bibfnamefont {S.~N.}\ \bibnamefont
  {Koptyaev}}, \bibinfo {author} {\bibfnamefont {A.~S.}\ \bibnamefont
  {Voloshin}}, \bibinfo {author} {\bibfnamefont {N.~M.}\ \bibnamefont
  {Kondratiev}}, \bibinfo {author} {\bibfnamefont {K.~N.}\ \bibnamefont
  {Min'kov}}, \bibinfo {author} {\bibfnamefont {V.~E.}\ \bibnamefont
  {Lobanov}}, \bibinfo {author} {\bibfnamefont {M.~V.}\ \bibnamefont {Ryabko}},
  \bibinfo {author} {\bibfnamefont {S.~V.}\ \bibnamefont {Polonsky}}, \ and\
  \bibinfo {author} {\bibfnamefont {I.~A.}\ \bibnamefont {Bilenko}},\ }\href
  {\doibase 10.1103/PhysRevApplied.18.034068} {\bibfield  {journal} {\bibinfo
  {journal} {Phys. Rev. Applied}\ }\textbf {\bibinfo {volume} {18}},\ \bibinfo
  {pages} {034068} (\bibinfo {year} {2022})}\BibitemShut {NoStop}%
\bibitem [{\citenamefont {Jang}\ \emph {et~al.}(2021)\citenamefont {Jang},
  \citenamefont {Okawachi}, \citenamefont {Zhao}, \citenamefont {Ji},
  \citenamefont {Joshi}, \citenamefont {Lipson},\ and\ \citenamefont
  {Gaeta}}]{jang2021conversion}%
  \BibitemOpen
  \bibfield  {author} {\bibinfo {author} {\bibfnamefont {J.~K.}\ \bibnamefont
  {Jang}}, \bibinfo {author} {\bibfnamefont {Y.}~\bibnamefont {Okawachi}},
  \bibinfo {author} {\bibfnamefont {Y.}~\bibnamefont {Zhao}}, \bibinfo {author}
  {\bibfnamefont {X.}~\bibnamefont {Ji}}, \bibinfo {author} {\bibfnamefont
  {C.}~\bibnamefont {Joshi}}, \bibinfo {author} {\bibfnamefont
  {M.}~\bibnamefont {Lipson}}, \ and\ \bibinfo {author} {\bibfnamefont {A.~L.}\
  \bibnamefont {Gaeta}},\ }\href {\doibase 10.1364/OL.423654} {\bibfield
  {journal} {\bibinfo  {journal} {Opt. Lett.}\ }\textbf {\bibinfo {volume}
  {46}},\ \bibinfo {pages} {3657} (\bibinfo {year} {2021})}\BibitemShut
  {NoStop}%
\bibitem [{\citenamefont {Bao}\ \emph {et~al.}(2014)\citenamefont {Bao},
  \citenamefont {Zhang}, \citenamefont {Matsko}, \citenamefont {Yan},
  \citenamefont {Zhao}, \citenamefont {Xie}, \citenamefont {Agarwal},
  \citenamefont {Kimerling}, \citenamefont {Michel}, \citenamefont {Maleki},\
  and\ \citenamefont {Willner}}]{Bao:14}%
  \BibitemOpen
  \bibfield  {author} {\bibinfo {author} {\bibfnamefont {C.}~\bibnamefont
  {Bao}}, \bibinfo {author} {\bibfnamefont {L.}~\bibnamefont {Zhang}}, \bibinfo
  {author} {\bibfnamefont {A.}~\bibnamefont {Matsko}}, \bibinfo {author}
  {\bibfnamefont {Y.}~\bibnamefont {Yan}}, \bibinfo {author} {\bibfnamefont
  {Z.}~\bibnamefont {Zhao}}, \bibinfo {author} {\bibfnamefont {G.}~\bibnamefont
  {Xie}}, \bibinfo {author} {\bibfnamefont {A.~M.}\ \bibnamefont {Agarwal}},
  \bibinfo {author} {\bibfnamefont {L.~C.}\ \bibnamefont {Kimerling}}, \bibinfo
  {author} {\bibfnamefont {J.}~\bibnamefont {Michel}}, \bibinfo {author}
  {\bibfnamefont {L.}~\bibnamefont {Maleki}}, \ and\ \bibinfo {author}
  {\bibfnamefont {A.~E.}\ \bibnamefont {Willner}},\ }\href {\doibase
  10.1364/OL.39.006126} {\bibfield  {journal} {\bibinfo  {journal} {Opt.
  Lett.}\ }\textbf {\bibinfo {volume} {39}},\ \bibinfo {pages} {6126} (\bibinfo
  {year} {2014})}\BibitemShut {NoStop}%
\bibitem [{\citenamefont {G\"artner}\ \emph {et~al.}(2019)\citenamefont
  {G\"artner}, \citenamefont {Trocha}, \citenamefont {Mandel}, \citenamefont
  {Koos}, \citenamefont {Jahnke},\ and\ \citenamefont {Reichel}}]{Gartner2019}%
  \BibitemOpen
  \bibfield  {author} {\bibinfo {author} {\bibfnamefont {J.}~\bibnamefont
  {G\"artner}}, \bibinfo {author} {\bibfnamefont {P.}~\bibnamefont {Trocha}},
  \bibinfo {author} {\bibfnamefont {R.}~\bibnamefont {Mandel}}, \bibinfo
  {author} {\bibfnamefont {C.}~\bibnamefont {Koos}}, \bibinfo {author}
  {\bibfnamefont {T.}~\bibnamefont {Jahnke}}, \ and\ \bibinfo {author}
  {\bibfnamefont {W.}~\bibnamefont {Reichel}},\ }\href {\doibase
  10.1103/PhysRevA.100.033819} {\bibfield  {journal} {\bibinfo  {journal}
  {Phys. Rev. A}\ }\textbf {\bibinfo {volume} {100}},\ \bibinfo {pages}
  {033819} (\bibinfo {year} {2019})}\BibitemShut {NoStop}%
\bibitem [{\citenamefont {Herr}\ \emph {et~al.}(2014)\citenamefont {Herr},
  \citenamefont {Brasch}, \citenamefont {Jost}, \citenamefont {Wang},
  \citenamefont {Kondratiev}, \citenamefont {Gorodetsky},\ and\ \citenamefont
  {Kippenberg}}]{herr2014temporal}%
  \BibitemOpen
  \bibfield  {author} {\bibinfo {author} {\bibfnamefont {T.}~\bibnamefont
  {Herr}}, \bibinfo {author} {\bibfnamefont {V.}~\bibnamefont {Brasch}},
  \bibinfo {author} {\bibfnamefont {J.~D.}\ \bibnamefont {Jost}}, \bibinfo
  {author} {\bibfnamefont {C.~Y.}\ \bibnamefont {Wang}}, \bibinfo {author}
  {\bibfnamefont {N.~M.}\ \bibnamefont {Kondratiev}}, \bibinfo {author}
  {\bibfnamefont {M.~L.}\ \bibnamefont {Gorodetsky}}, \ and\ \bibinfo {author}
  {\bibfnamefont {T.~J.}\ \bibnamefont {Kippenberg}},\ }\href@noop {}
  {\bibfield  {journal} {\bibinfo  {journal} {Nat. Photon.}\ }\textbf {\bibinfo
  {volume} {8}},\ \bibinfo {pages} {145} (\bibinfo {year} {2014})}\BibitemShut
  {NoStop}%
\bibitem [{\citenamefont {Pavlov}\ \emph {et~al.}(2018)\citenamefont {Pavlov},
  \citenamefont {Koptyaev}, \citenamefont {Lihachev}, \citenamefont {Voloshin},
  \citenamefont {Gorodnitskiy}, \citenamefont {Ryabko}, \citenamefont
  {Polonsky},\ and\ \citenamefont {Gorodetsky}}]{Pavlov2018}%
  \BibitemOpen
  \bibfield  {author} {\bibinfo {author} {\bibfnamefont {N.~G.}\ \bibnamefont
  {Pavlov}}, \bibinfo {author} {\bibfnamefont {S.}~\bibnamefont {Koptyaev}},
  \bibinfo {author} {\bibfnamefont {G.~V.}\ \bibnamefont {Lihachev}}, \bibinfo
  {author} {\bibfnamefont {A.~S.}\ \bibnamefont {Voloshin}}, \bibinfo {author}
  {\bibfnamefont {A.~S.}\ \bibnamefont {Gorodnitskiy}}, \bibinfo {author}
  {\bibfnamefont {M.~V.}\ \bibnamefont {Ryabko}}, \bibinfo {author}
  {\bibfnamefont {S.~V.}\ \bibnamefont {Polonsky}}, \ and\ \bibinfo {author}
  {\bibfnamefont {M.~L.}\ \bibnamefont {Gorodetsky}},\ }\href {\doibase
  10.1038/s41566-018-0277-2} {\bibfield  {journal} {\bibinfo  {journal} {Nature
  Photonics}\ }\textbf {\bibinfo {volume} {12}},\ \bibinfo {pages} {694}
  (\bibinfo {year} {2018})}\BibitemShut {NoStop}%
\bibitem [{\citenamefont {Raja}\ \emph {et~al.}(2019)\citenamefont {Raja},
  \citenamefont {Voloshin}, \citenamefont {Guo}, \citenamefont {Agafonova},
  \citenamefont {Liu}, \citenamefont {Gorodnitskiy}, \citenamefont {Karpov},
  \citenamefont {Pavlov}, \citenamefont {Lucas}, \citenamefont {Galiev},
  \citenamefont {Shitikov}, \citenamefont {Jost}, \citenamefont {Gorodetsky},\
  and\ \citenamefont {Kippenberg}}]{Raja2019}%
  \BibitemOpen
  \bibfield  {author} {\bibinfo {author} {\bibfnamefont {A.~S.}\ \bibnamefont
  {Raja}}, \bibinfo {author} {\bibfnamefont {A.~S.}\ \bibnamefont {Voloshin}},
  \bibinfo {author} {\bibfnamefont {H.}~\bibnamefont {Guo}}, \bibinfo {author}
  {\bibfnamefont {S.~E.}\ \bibnamefont {Agafonova}}, \bibinfo {author}
  {\bibfnamefont {J.}~\bibnamefont {Liu}}, \bibinfo {author} {\bibfnamefont
  {A.~S.}\ \bibnamefont {Gorodnitskiy}}, \bibinfo {author} {\bibfnamefont
  {M.}~\bibnamefont {Karpov}}, \bibinfo {author} {\bibfnamefont {N.~G.}\
  \bibnamefont {Pavlov}}, \bibinfo {author} {\bibfnamefont {E.}~\bibnamefont
  {Lucas}}, \bibinfo {author} {\bibfnamefont {R.~R.}\ \bibnamefont {Galiev}},
  \bibinfo {author} {\bibfnamefont {A.~E.}\ \bibnamefont {Shitikov}}, \bibinfo
  {author} {\bibfnamefont {J.~D.}\ \bibnamefont {Jost}}, \bibinfo {author}
  {\bibfnamefont {M.~L.}\ \bibnamefont {Gorodetsky}}, \ and\ \bibinfo {author}
  {\bibfnamefont {T.~J.}\ \bibnamefont {Kippenberg}},\ }\href {\doibase
  10.1038/s41467-019-08498-2} {\bibfield  {journal} {\bibinfo  {journal}
  {Nature Communications}\ }\textbf {\bibinfo {volume} {10}},\ \bibinfo {pages}
  {680} (\bibinfo {year} {2019})}\BibitemShut {NoStop}%
\bibitem [{\citenamefont {Shen}\ \emph {et~al.}(2020)\citenamefont {Shen},
  \citenamefont {Chang}, \citenamefont {Liu}, \citenamefont {Wang},
  \citenamefont {Yang}, \citenamefont {Xiang}, \citenamefont {Wang},
  \citenamefont {He}, \citenamefont {Liu}, \citenamefont {Xie}, \citenamefont
  {Guo}, \citenamefont {Kinghorn}, \citenamefont {Wu}, \citenamefont {Ji},
  \citenamefont {Kippenberg}, \citenamefont {Vahala},\ and\ \citenamefont
  {Bowers}}]{shen2019}%
  \BibitemOpen
  \bibfield  {author} {\bibinfo {author} {\bibfnamefont {B.}~\bibnamefont
  {Shen}}, \bibinfo {author} {\bibfnamefont {L.}~\bibnamefont {Chang}},
  \bibinfo {author} {\bibfnamefont {J.}~\bibnamefont {Liu}}, \bibinfo {author}
  {\bibfnamefont {H.}~\bibnamefont {Wang}}, \bibinfo {author} {\bibfnamefont
  {Q.-F.}\ \bibnamefont {Yang}}, \bibinfo {author} {\bibfnamefont
  {C.}~\bibnamefont {Xiang}}, \bibinfo {author} {\bibfnamefont {R.~N.}\
  \bibnamefont {Wang}}, \bibinfo {author} {\bibfnamefont {J.}~\bibnamefont
  {He}}, \bibinfo {author} {\bibfnamefont {T.}~\bibnamefont {Liu}}, \bibinfo
  {author} {\bibfnamefont {W.}~\bibnamefont {Xie}}, \bibinfo {author}
  {\bibfnamefont {J.}~\bibnamefont {Guo}}, \bibinfo {author} {\bibfnamefont
  {D.}~\bibnamefont {Kinghorn}}, \bibinfo {author} {\bibfnamefont
  {L.}~\bibnamefont {Wu}}, \bibinfo {author} {\bibfnamefont {Q.-X.}\
  \bibnamefont {Ji}}, \bibinfo {author} {\bibfnamefont {T.~J.}\ \bibnamefont
  {Kippenberg}}, \bibinfo {author} {\bibfnamefont {K.}~\bibnamefont {Vahala}},
  \ and\ \bibinfo {author} {\bibfnamefont {J.~E.}\ \bibnamefont {Bowers}},\
  }\href {\doibase 10.1038/s41586-020-2358-x} {\bibfield  {journal} {\bibinfo
  {journal} {Nature}\ }\textbf {\bibinfo {volume} {582}},\ \bibinfo {pages}
  {365} (\bibinfo {year} {2020})}\BibitemShut {NoStop}%
\bibitem [{\citenamefont {Yi}\ \emph {et~al.}(2016)\citenamefont {Yi},
  \citenamefont {Yang}, \citenamefont {Yang},\ and\ \citenamefont
  {Vahala}}]{Yi:16}%
  \BibitemOpen
  \bibfield  {author} {\bibinfo {author} {\bibfnamefont {X.}~\bibnamefont
  {Yi}}, \bibinfo {author} {\bibfnamefont {Q.-F.}\ \bibnamefont {Yang}},
  \bibinfo {author} {\bibfnamefont {K.~Y.}\ \bibnamefont {Yang}}, \ and\
  \bibinfo {author} {\bibfnamefont {K.}~\bibnamefont {Vahala}},\ }\href
  {\doibase 10.1364/OL.41.003419} {\bibfield  {journal} {\bibinfo  {journal}
  {Opt. Lett.}\ }\textbf {\bibinfo {volume} {41}},\ \bibinfo {pages} {3419}
  (\bibinfo {year} {2016})}\BibitemShut {NoStop}%
\bibitem [{\citenamefont {Gorodetsky}\ and\ \citenamefont
  {Ilchenko}(1999)}]{Gorodetsky:99}%
  \BibitemOpen
  \bibfield  {author} {\bibinfo {author} {\bibfnamefont {M.~L.}\ \bibnamefont
  {Gorodetsky}}\ and\ \bibinfo {author} {\bibfnamefont {V.~S.}\ \bibnamefont
  {Ilchenko}},\ }\href {\doibase 10.1364/JOSAB.16.000147} {\bibfield  {journal}
  {\bibinfo  {journal} {J. Opt. Soc. Am. B}\ }\textbf {\bibinfo {volume}
  {16}},\ \bibinfo {pages} {147} (\bibinfo {year} {1999})}\BibitemShut
  {NoStop}%
\bibitem [{\citenamefont {Galiev}\ \emph {et~al.}(2020)\citenamefont {Galiev},
  \citenamefont {Kondratiev}, \citenamefont {Lobanov}, \citenamefont {Matsko},\
  and\ \citenamefont {Bilenko}}]{Galiev2020}%
  \BibitemOpen
  \bibfield  {author} {\bibinfo {author} {\bibfnamefont {R.~R.}\ \bibnamefont
  {Galiev}}, \bibinfo {author} {\bibfnamefont {N.~M.}\ \bibnamefont
  {Kondratiev}}, \bibinfo {author} {\bibfnamefont {V.~E.}\ \bibnamefont
  {Lobanov}}, \bibinfo {author} {\bibfnamefont {A.~B.}\ \bibnamefont {Matsko}},
  \ and\ \bibinfo {author} {\bibfnamefont {I.~A.}\ \bibnamefont {Bilenko}},\
  }\href {\doibase 10.1103/PhysRevApplied.14.014036} {\bibfield  {journal}
  {\bibinfo  {journal} {Phys. Rev. Applied}\ }\textbf {\bibinfo {volume}
  {14}},\ \bibinfo {pages} {014036} (\bibinfo {year} {2020})}\BibitemShut
  {NoStop}%
\bibitem [{\citenamefont {Kondratiev}\ \emph {et~al.}(2017)\citenamefont
  {Kondratiev}, \citenamefont {Lobanov}, \citenamefont {Cherenkov},
  \citenamefont {Voloshin}, \citenamefont {Pavlov}, \citenamefont {Koptyaev},\
  and\ \citenamefont {Gorodetsky}}]{Kondratiev:17}%
  \BibitemOpen
  \bibfield  {author} {\bibinfo {author} {\bibfnamefont {N.~M.}\ \bibnamefont
  {Kondratiev}}, \bibinfo {author} {\bibfnamefont {V.~E.}\ \bibnamefont
  {Lobanov}}, \bibinfo {author} {\bibfnamefont {A.~V.}\ \bibnamefont
  {Cherenkov}}, \bibinfo {author} {\bibfnamefont {A.~S.}\ \bibnamefont
  {Voloshin}}, \bibinfo {author} {\bibfnamefont {N.~G.}\ \bibnamefont
  {Pavlov}}, \bibinfo {author} {\bibfnamefont {S.}~\bibnamefont {Koptyaev}}, \
  and\ \bibinfo {author} {\bibfnamefont {M.~L.}\ \bibnamefont {Gorodetsky}},\
  }\href {\doibase 10.1364/OE.25.028167} {\bibfield  {journal} {\bibinfo
  {journal} {Opt. Express}\ }\textbf {\bibinfo {volume} {25}},\ \bibinfo
  {pages} {28167} (\bibinfo {year} {2017})}\BibitemShut {NoStop}%
\bibitem [{\citenamefont {Kondratiev}\ \emph {et~al.}(2023)\citenamefont
  {Kondratiev}, \citenamefont {Lobanov}, \citenamefont {Shitikov},
  \citenamefont {Galiev}, \citenamefont {Chermoshentsev}, \citenamefont
  {Dmitriev}, \citenamefont {Danilin}, \citenamefont {Lonshakov}, \citenamefont
  {Min’kov}, \citenamefont {Sokol}, \citenamefont {Cordette}, \citenamefont
  {Luo}, \citenamefont {Liang}, \citenamefont {Liu},\ and\ \citenamefont
  {Bilenko}}]{Kondratyev:22}%
  \BibitemOpen
  \bibfield  {author} {\bibinfo {author} {\bibfnamefont {N.~M.}\ \bibnamefont
  {Kondratiev}}, \bibinfo {author} {\bibfnamefont {V.~E.}\ \bibnamefont
  {Lobanov}}, \bibinfo {author} {\bibfnamefont {A.~E.}\ \bibnamefont
  {Shitikov}}, \bibinfo {author} {\bibfnamefont {R.~R.}\ \bibnamefont
  {Galiev}}, \bibinfo {author} {\bibfnamefont {D.~A.}\ \bibnamefont
  {Chermoshentsev}}, \bibinfo {author} {\bibfnamefont {N.~Y.}\ \bibnamefont
  {Dmitriev}}, \bibinfo {author} {\bibfnamefont {A.~N.}\ \bibnamefont
  {Danilin}}, \bibinfo {author} {\bibfnamefont {E.~A.}\ \bibnamefont
  {Lonshakov}}, \bibinfo {author} {\bibfnamefont {K.~N.}\ \bibnamefont
  {Min’kov}}, \bibinfo {author} {\bibfnamefont {D.~M.}\ \bibnamefont
  {Sokol}}, \bibinfo {author} {\bibfnamefont {S.~J.}\ \bibnamefont {Cordette}},
  \bibinfo {author} {\bibfnamefont {Y.-H.}\ \bibnamefont {Luo}}, \bibinfo
  {author} {\bibfnamefont {W.}~\bibnamefont {Liang}}, \bibinfo {author}
  {\bibfnamefont {J.}~\bibnamefont {Liu}}, \ and\ \bibinfo {author}
  {\bibfnamefont {I.~A.}\ \bibnamefont {Bilenko}},\ }\href {\doibase
  10.1007/s11467-022-1245-3} {\bibfield  {journal} {\bibinfo  {journal}
  {Frontiers of Physics}\ }\textbf {\bibinfo {volume} {18}},\ \bibinfo {eid}
  {21305} (\bibinfo {year} {2023})}\BibitemShut {NoStop}%
\bibitem [{\citenamefont {Grudinin}\ and\ \citenamefont
  {Yu}(2015)}]{Grudinin:15}%
  \BibitemOpen
  \bibfield  {author} {\bibinfo {author} {\bibfnamefont {I.~S.}\ \bibnamefont
  {Grudinin}}\ and\ \bibinfo {author} {\bibfnamefont {N.}~\bibnamefont {Yu}},\
  }\href {\doibase 10.1364/OPTICA.2.000221} {\bibfield  {journal} {\bibinfo
  {journal} {Optica}\ }\textbf {\bibinfo {volume} {2}},\ \bibinfo {pages} {221}
  (\bibinfo {year} {2015})}\BibitemShut {NoStop}%
\bibitem [{\citenamefont {Fujii}\ and\ \citenamefont
  {Tanabe}(2020)}]{FujiiTanabe2020}%
  \BibitemOpen
  \bibfield  {author} {\bibinfo {author} {\bibfnamefont {S.}~\bibnamefont
  {Fujii}}\ and\ \bibinfo {author} {\bibfnamefont {T.}~\bibnamefont {Tanabe}},\
  }\href {\doibase doi:10.1515/nanoph-2019-0497} {\bibfield  {journal}
  {\bibinfo  {journal} {Nanophotonics}\ }\textbf {\bibinfo {volume} {9}},\
  \bibinfo {pages} {1087} (\bibinfo {year} {2020})}\BibitemShut {NoStop}%
\bibitem [{\citenamefont {Wang}\ \emph {et~al.}(2022)\citenamefont {Wang},
  \citenamefont {Lee}, \citenamefont {Chen},\ and\ \citenamefont
  {Wang}}]{Wang:22}%
  \BibitemOpen
  \bibfield  {author} {\bibinfo {author} {\bibfnamefont {S.-P.}\ \bibnamefont
  {Wang}}, \bibinfo {author} {\bibfnamefont {T.-H.}\ \bibnamefont {Lee}},
  \bibinfo {author} {\bibfnamefont {Y.-Y.}\ \bibnamefont {Chen}}, \ and\
  \bibinfo {author} {\bibfnamefont {P.-H.}\ \bibnamefont {Wang}},\ }\href
  {\doibase 10.3390/mi13030454} {\bibfield  {journal} {\bibinfo  {journal}
  {Micromachines}\ }\textbf {\bibinfo {volume} {13}} (\bibinfo {year} {2022}),\
  10.3390/mi13030454}\BibitemShut {NoStop}%
\bibitem [{\citenamefont {Lucas}\ \emph {et~al.}(2022)\citenamefont {Lucas},
  \citenamefont {Yu}, \citenamefont {Briles}, \citenamefont {Carlson},\ and\
  \citenamefont {Papp}}]{Lucas:22_arx}%
  \BibitemOpen
  \bibfield  {author} {\bibinfo {author} {\bibfnamefont {E.}~\bibnamefont
  {Lucas}}, \bibinfo {author} {\bibfnamefont {S.-P.}\ \bibnamefont {Yu}},
  \bibinfo {author} {\bibfnamefont {T.~C.}\ \bibnamefont {Briles}}, \bibinfo
  {author} {\bibfnamefont {D.~R.}\ \bibnamefont {Carlson}}, \ and\ \bibinfo
  {author} {\bibfnamefont {S.~B.}\ \bibnamefont {Papp}},\ }\href {\doibase
  10.48550/ARXIV.2209.10294} {\enquote {\bibinfo {title} {Tailoring microcombs
  with inverse-designed, meta-dispersion microresonators},}\ } (\bibinfo {year}
  {2022}),\ \Eprint {http://arxiv.org/abs/2209.10294} {arXiv:2209.10294
  [physics.optics]} \BibitemShut {NoStop}%
\bibitem [{\citenamefont {Zhang}\ \emph {et~al.}(2022)\citenamefont {Zhang},
  \citenamefont {Kang}, \citenamefont {Wang}, \citenamefont {Pan},\ and\
  \citenamefont {Qu}}]{Zhang:22}%
  \BibitemOpen
  \bibfield  {author} {\bibinfo {author} {\bibfnamefont {C.}~\bibnamefont
  {Zhang}}, \bibinfo {author} {\bibfnamefont {G.}~\bibnamefont {Kang}},
  \bibinfo {author} {\bibfnamefont {J.}~\bibnamefont {Wang}}, \bibinfo {author}
  {\bibfnamefont {Y.}~\bibnamefont {Pan}}, \ and\ \bibinfo {author}
  {\bibfnamefont {J.}~\bibnamefont {Qu}},\ }\href {\doibase 10.1364/OE.471706}
  {\bibfield  {journal} {\bibinfo  {journal} {Opt. Express}\ }\textbf {\bibinfo
  {volume} {30}},\ \bibinfo {pages} {44395} (\bibinfo {year}
  {2022})}\BibitemShut {NoStop}%
\bibitem [{\citenamefont {Chembo}\ and\ \citenamefont
  {Menyuk}(2013)}]{Chembo2013}%
  \BibitemOpen
  \bibfield  {author} {\bibinfo {author} {\bibfnamefont {Y.~K.}\ \bibnamefont
  {Chembo}}\ and\ \bibinfo {author} {\bibfnamefont {C.~R.}\ \bibnamefont
  {Menyuk}},\ }\href {\doibase 10.1103/PhysRevA.87.053852} {\bibfield
  {journal} {\bibinfo  {journal} {Phys. Rev. A}\ }\textbf {\bibinfo {volume}
  {87}},\ \bibinfo {pages} {053852} (\bibinfo {year} {2013})}\BibitemShut
  {NoStop}%
\bibitem [{\citenamefont {Godey}\ \emph {et~al.}(2014)\citenamefont {Godey},
  \citenamefont {Balakireva}, \citenamefont {Coillet},\ and\ \citenamefont
  {Chembo}}]{Godey2014}%
  \BibitemOpen
  \bibfield  {author} {\bibinfo {author} {\bibfnamefont {C.}~\bibnamefont
  {Godey}}, \bibinfo {author} {\bibfnamefont {I.~V.}\ \bibnamefont
  {Balakireva}}, \bibinfo {author} {\bibfnamefont {A.}~\bibnamefont {Coillet}},
  \ and\ \bibinfo {author} {\bibfnamefont {Y.~K.}\ \bibnamefont {Chembo}},\
  }\href {\doibase 10.1103/PhysRevA.89.063814} {\bibfield  {journal} {\bibinfo
  {journal} {Phys. Rev. A}\ }\textbf {\bibinfo {volume} {89}},\ \bibinfo
  {pages} {063814} (\bibinfo {year} {2014})}\BibitemShut {NoStop}%
\bibitem [{\citenamefont {Bao}\ \emph {et~al.}(2017)\citenamefont {Bao},
  \citenamefont {Taheri}, \citenamefont {Zhang}, \citenamefont {Matsko},
  \citenamefont {Yan}, \citenamefont {Liao}, \citenamefont {Maleki},\ and\
  \citenamefont {Willner}}]{Bao:17_d3}%
  \BibitemOpen
  \bibfield  {author} {\bibinfo {author} {\bibfnamefont {C.}~\bibnamefont
  {Bao}}, \bibinfo {author} {\bibfnamefont {H.}~\bibnamefont {Taheri}},
  \bibinfo {author} {\bibfnamefont {L.}~\bibnamefont {Zhang}}, \bibinfo
  {author} {\bibfnamefont {A.}~\bibnamefont {Matsko}}, \bibinfo {author}
  {\bibfnamefont {Y.}~\bibnamefont {Yan}}, \bibinfo {author} {\bibfnamefont
  {P.}~\bibnamefont {Liao}}, \bibinfo {author} {\bibfnamefont {L.}~\bibnamefont
  {Maleki}}, \ and\ \bibinfo {author} {\bibfnamefont {A.~E.}\ \bibnamefont
  {Willner}},\ }\href {\doibase 10.1364/JOSAB.34.000715} {\bibfield  {journal}
  {\bibinfo  {journal} {J. Opt. Soc. Am. B}\ }\textbf {\bibinfo {volume}
  {34}},\ \bibinfo {pages} {715} (\bibinfo {year} {2017})}\BibitemShut
  {NoStop}%
\bibitem [{\citenamefont {Mb\'{e}}\ and\ \citenamefont
  {Chembo}(2020)}]{TallaMbe:20}%
  \BibitemOpen
  \bibfield  {author} {\bibinfo {author} {\bibfnamefont {J.~H.~T.}\
  \bibnamefont {Mb\'{e}}}\ and\ \bibinfo {author} {\bibfnamefont {Y.~K.}\
  \bibnamefont {Chembo}},\ }\href {\doibase 10.1364/JOSAB.396610} {\bibfield
  {journal} {\bibinfo  {journal} {J. Opt. Soc. Am. B}\ }\textbf {\bibinfo
  {volume} {37}},\ \bibinfo {pages} {A69} (\bibinfo {year} {2020})}\BibitemShut
  {NoStop}%
\bibitem [{\citenamefont {Anderson}\ \emph {et~al.}(2022)\citenamefont
  {Anderson}, \citenamefont {Weng}, \citenamefont {Lihachev}, \citenamefont
  {Tikan}, \citenamefont {Liu},\ and\ \citenamefont
  {Kippenberg}}]{Anderson2022}%
  \BibitemOpen
  \bibfield  {author} {\bibinfo {author} {\bibfnamefont {M.~H.}\ \bibnamefont
  {Anderson}}, \bibinfo {author} {\bibfnamefont {W.}~\bibnamefont {Weng}},
  \bibinfo {author} {\bibfnamefont {G.}~\bibnamefont {Lihachev}}, \bibinfo
  {author} {\bibfnamefont {A.}~\bibnamefont {Tikan}}, \bibinfo {author}
  {\bibfnamefont {J.}~\bibnamefont {Liu}}, \ and\ \bibinfo {author}
  {\bibfnamefont {T.~J.}\ \bibnamefont {Kippenberg}},\ }\href {\doibase
  10.1038/s41467-022-31916-x} {\bibfield  {journal} {\bibinfo  {journal}
  {Nature Communications}\ }\textbf {\bibinfo {volume} {13}},\ \bibinfo {pages}
  {4764} (\bibinfo {year} {2022})}\BibitemShut {NoStop}%
\bibitem [{\citenamefont {Xiao}\ \emph {et~al.}(2023)\citenamefont {Xiao},
  \citenamefont {Li}, \citenamefont {Cai}, \citenamefont {Zhang}, \citenamefont
  {Huang}, \citenamefont {Li}, \citenamefont {Yao}, \citenamefont {Wu},\ and\
  \citenamefont {Chen}}]{Xiao2023}%
  \BibitemOpen
  \bibfield  {author} {\bibinfo {author} {\bibfnamefont {Z.}~\bibnamefont
  {Xiao}}, \bibinfo {author} {\bibfnamefont {T.}~\bibnamefont {Li}}, \bibinfo
  {author} {\bibfnamefont {M.}~\bibnamefont {Cai}}, \bibinfo {author}
  {\bibfnamefont {H.}~\bibnamefont {Zhang}}, \bibinfo {author} {\bibfnamefont
  {Y.}~\bibnamefont {Huang}}, \bibinfo {author} {\bibfnamefont
  {C.}~\bibnamefont {Li}}, \bibinfo {author} {\bibfnamefont {B.}~\bibnamefont
  {Yao}}, \bibinfo {author} {\bibfnamefont {K.}~\bibnamefont {Wu}}, \ and\
  \bibinfo {author} {\bibfnamefont {J.}~\bibnamefont {Chen}},\ }\href {\doibase
  10.1038/s41377-023-01076-8} {\bibfield  {journal} {\bibinfo  {journal}
  {Light: Science {\&} Applications}\ }\textbf {\bibinfo {volume} {12}},\
  \bibinfo {pages} {33} (\bibinfo {year} {2023})}\BibitemShut {NoStop}%
\bibitem [{\citenamefont {Zhang}\ \emph {et~al.}(2023)\citenamefont {Zhang},
  \citenamefont {Bi},\ and\ \citenamefont
  {Del'Haye}}]{zhang2022microresonator}%
  \BibitemOpen
  \bibfield  {author} {\bibinfo {author} {\bibfnamefont {S.}~\bibnamefont
  {Zhang}}, \bibinfo {author} {\bibfnamefont {T.}~\bibnamefont {Bi}}, \ and\
  \bibinfo {author} {\bibfnamefont {P.}~\bibnamefont {Del'Haye}},\ }\href
  {\doibase 10.1002/lpor.202300075} {\bibfield  {journal} {\bibinfo  {journal}
  {Laser \& Photonics Reviews}\ ,\ \bibinfo {pages} {2300075}} (\bibinfo {year}
  {2023})}\BibitemShut {NoStop}%
\bibitem [{\citenamefont {Guo}\ \emph {et~al.}(2017)\citenamefont {Guo},
  \citenamefont {Karpov}, \citenamefont {Lucas}, \citenamefont {Kordts},
  \citenamefont {Pfeiffer}, \citenamefont {Brasch}, \citenamefont {Lihachev},
  \citenamefont {Lobanov}, \citenamefont {Gorodetsky},\ and\ \citenamefont
  {Kippenberg}}]{Guo2017}%
  \BibitemOpen
  \bibfield  {author} {\bibinfo {author} {\bibfnamefont {H.}~\bibnamefont
  {Guo}}, \bibinfo {author} {\bibfnamefont {M.}~\bibnamefont {Karpov}},
  \bibinfo {author} {\bibfnamefont {E.}~\bibnamefont {Lucas}}, \bibinfo
  {author} {\bibfnamefont {A.}~\bibnamefont {Kordts}}, \bibinfo {author}
  {\bibfnamefont {M.~H.~P.}\ \bibnamefont {Pfeiffer}}, \bibinfo {author}
  {\bibfnamefont {V.}~\bibnamefont {Brasch}}, \bibinfo {author} {\bibfnamefont
  {G.}~\bibnamefont {Lihachev}}, \bibinfo {author} {\bibfnamefont {V.~E.}\
  \bibnamefont {Lobanov}}, \bibinfo {author} {\bibfnamefont {M.~L.}\
  \bibnamefont {Gorodetsky}}, \ and\ \bibinfo {author} {\bibfnamefont {T.~J.}\
  \bibnamefont {Kippenberg}},\ }\href {\doibase 10.1038/nphys3893} {\bibfield
  {journal} {\bibinfo  {journal} {Nature Physics}\ }\textbf {\bibinfo {volume}
  {13}},\ \bibinfo {pages} {94} (\bibinfo {year} {2017})}\BibitemShut {NoStop}%
\bibitem [{\citenamefont {Li}\ \emph {et~al.}(2018)\citenamefont {Li},
  \citenamefont {Shen}, \citenamefont {Wang}, \citenamefont {Yang},
  \citenamefont {Yi}, \citenamefont {Yang}, \citenamefont {Zhou},\ and\
  \citenamefont {Vahala}}]{Li:18}%
  \BibitemOpen
  \bibfield  {author} {\bibinfo {author} {\bibfnamefont {X.}~\bibnamefont
  {Li}}, \bibinfo {author} {\bibfnamefont {B.}~\bibnamefont {Shen}}, \bibinfo
  {author} {\bibfnamefont {H.}~\bibnamefont {Wang}}, \bibinfo {author}
  {\bibfnamefont {K.~Y.}\ \bibnamefont {Yang}}, \bibinfo {author}
  {\bibfnamefont {X.}~\bibnamefont {Yi}}, \bibinfo {author} {\bibfnamefont
  {Q.-F.}\ \bibnamefont {Yang}}, \bibinfo {author} {\bibfnamefont
  {Z.}~\bibnamefont {Zhou}}, \ and\ \bibinfo {author} {\bibfnamefont
  {K.}~\bibnamefont {Vahala}},\ }\href {\doibase 10.1364/OL.43.002567}
  {\bibfield  {journal} {\bibinfo  {journal} {Opt. Lett.}\ }\textbf {\bibinfo
  {volume} {43}},\ \bibinfo {pages} {2567} (\bibinfo {year}
  {2018})}\BibitemShut {NoStop}%
\bibitem [{\citenamefont {Fujii}\ \emph {et~al.}(2017)\citenamefont {Fujii},
  \citenamefont {Hori}, \citenamefont {Kato}, \citenamefont {Suzuki},
  \citenamefont {Okabe}, \citenamefont {Yoshiki}, \citenamefont {Jinnai},\ and\
  \citenamefont {Tanabe}}]{Fujii:17}%
  \BibitemOpen
  \bibfield  {author} {\bibinfo {author} {\bibfnamefont {S.}~\bibnamefont
  {Fujii}}, \bibinfo {author} {\bibfnamefont {A.}~\bibnamefont {Hori}},
  \bibinfo {author} {\bibfnamefont {T.}~\bibnamefont {Kato}}, \bibinfo {author}
  {\bibfnamefont {R.}~\bibnamefont {Suzuki}}, \bibinfo {author} {\bibfnamefont
  {Y.}~\bibnamefont {Okabe}}, \bibinfo {author} {\bibfnamefont
  {W.}~\bibnamefont {Yoshiki}}, \bibinfo {author} {\bibfnamefont {A.-C.}\
  \bibnamefont {Jinnai}}, \ and\ \bibinfo {author} {\bibfnamefont
  {T.}~\bibnamefont {Tanabe}},\ }\href {\doibase 10.1364/OE.25.028969}
  {\bibfield  {journal} {\bibinfo  {journal} {Opt. Express}\ }\textbf {\bibinfo
  {volume} {25}},\ \bibinfo {pages} {28969} (\bibinfo {year}
  {2017})}\BibitemShut {NoStop}%
\bibitem [{\citenamefont {Yang}\ \emph {et~al.}(2017)\citenamefont {Yang},
  \citenamefont {Yi}, \citenamefont {Yang},\ and\ \citenamefont
  {Vahala}}]{Yang:2017}%
  \BibitemOpen
  \bibfield  {author} {\bibinfo {author} {\bibfnamefont {Q.-F.}\ \bibnamefont
  {Yang}}, \bibinfo {author} {\bibfnamefont {X.}~\bibnamefont {Yi}}, \bibinfo
  {author} {\bibfnamefont {K.~Y.}\ \bibnamefont {Yang}}, \ and\ \bibinfo
  {author} {\bibfnamefont {K.}~\bibnamefont {Vahala}},\ }\href {\doibase
  10.1038/nphoton.2017.117} {\bibfield  {journal} {\bibinfo  {journal} {Nature
  Photonics}\ }\textbf {\bibinfo {volume} {11}},\ \bibinfo {pages} {560}
  (\bibinfo {year} {2017})}\BibitemShut {NoStop}%
\bibitem [{\citenamefont {Lobanov}\ \emph {et~al.}(2020)\citenamefont
  {Lobanov}, \citenamefont {Shitikov}, \citenamefont {Galiev}, \citenamefont
  {Min'kov},\ and\ \citenamefont {Kondratiev}}]{Lobanov:20}%
  \BibitemOpen
  \bibfield  {author} {\bibinfo {author} {\bibfnamefont {V.~E.}\ \bibnamefont
  {Lobanov}}, \bibinfo {author} {\bibfnamefont {A.~E.}\ \bibnamefont
  {Shitikov}}, \bibinfo {author} {\bibfnamefont {R.~R.}\ \bibnamefont
  {Galiev}}, \bibinfo {author} {\bibfnamefont {K.~N.}\ \bibnamefont {Min'kov}},
  \ and\ \bibinfo {author} {\bibfnamefont {N.~M.}\ \bibnamefont {Kondratiev}},\
  }\href {\doibase 10.1364/OE.410318} {\bibfield  {journal} {\bibinfo
  {journal} {Opt. Express}\ }\textbf {\bibinfo {volume} {28}},\ \bibinfo
  {pages} {36544} (\bibinfo {year} {2020})}\BibitemShut {NoStop}%
\bibitem [{\citenamefont {Fan}\ and\ \citenamefont {Skryabin}(2020)}]{Fan:20}%
  \BibitemOpen
  \bibfield  {author} {\bibinfo {author} {\bibfnamefont {Z.}~\bibnamefont
  {Fan}}\ and\ \bibinfo {author} {\bibfnamefont {D.~V.}\ \bibnamefont
  {Skryabin}},\ }\href {\doibase 10.1364/OL.409362} {\bibfield  {journal}
  {\bibinfo  {journal} {Opt. Lett.}\ }\textbf {\bibinfo {volume} {45}},\
  \bibinfo {pages} {6446} (\bibinfo {year} {2020})}\BibitemShut {NoStop}%
\bibitem [{\citenamefont {Kondratiev}\ and\ \citenamefont
  {Lobanov}(2020)}]{Kondratiev:20}%
  \BibitemOpen
  \bibfield  {author} {\bibinfo {author} {\bibfnamefont {N.~M.}\ \bibnamefont
  {Kondratiev}}\ and\ \bibinfo {author} {\bibfnamefont {V.~E.}\ \bibnamefont
  {Lobanov}},\ }\href {\doibase 10.1103/PhysRevA.101.013816} {\bibfield
  {journal} {\bibinfo  {journal} {Phys. Rev. A}\ }\textbf {\bibinfo {volume}
  {101}},\ \bibinfo {pages} {013816} (\bibinfo {year} {2020})}\BibitemShut
  {NoStop}%
\bibitem [{\citenamefont {Skryabin}(2020)}]{Skryabin:20}%
  \BibitemOpen
  \bibfield  {author} {\bibinfo {author} {\bibfnamefont {D.~V.}\ \bibnamefont
  {Skryabin}},\ }\href {\doibase 10.1364/OSAC.392211} {\bibfield  {journal}
  {\bibinfo  {journal} {OSA Continuum}\ }\textbf {\bibinfo {volume} {3}},\
  \bibinfo {pages} {1364} (\bibinfo {year} {2020})}\BibitemShut {NoStop}%
\bibitem [{\citenamefont {Ulanov}\ \emph {et~al.}(2023)\citenamefont {Ulanov},
  \citenamefont {Wildi}, \citenamefont {Pavlov}, \citenamefont {Jost},
  \citenamefont {Karpov},\ and\ \citenamefont {Herr}}]{Ulanov:22}%
  \BibitemOpen
  \bibfield  {author} {\bibinfo {author} {\bibfnamefont {A.~E.}\ \bibnamefont
  {Ulanov}}, \bibinfo {author} {\bibfnamefont {T.}~\bibnamefont {Wildi}},
  \bibinfo {author} {\bibfnamefont {N.~G.}\ \bibnamefont {Pavlov}}, \bibinfo
  {author} {\bibfnamefont {J.~D.}\ \bibnamefont {Jost}}, \bibinfo {author}
  {\bibfnamefont {M.}~\bibnamefont {Karpov}}, \ and\ \bibinfo {author}
  {\bibfnamefont {T.}~\bibnamefont {Herr}},\ }\href {\doibase
  10.48550/ARXIV.2301.13132} {\enquote {\bibinfo {title} {Synthetic-reflection
  self-injection-locked microcombs},}\ } (\bibinfo {year} {2023}),\ \Eprint
  {http://arxiv.org/abs/2301.13132} {arXiv:2301.13132 [physics.optics]}
  \BibitemShut {NoStop}%
\end{thebibliography}%

\end{document}